\documentclass[twocolumn]{aastex63}
\usepackage{CJK}
\usepackage{graphics}

\def\cm-3{\,{\rm cm^{-3}}}

\def\kpc-3{\,{\rm kpc^{-3}}}
\def\myr-1{\,{\rm Myr^{-1}}}

\def\kpc{\,{\rm kpc}}

\def\t0{$t_{\rm{cool,0}}\ $}
\def\rsn{$R_{\rm{SN}}\ $}
\def\tmulti{$t_{\rm{multi}}\ $}

\def\vrms{$v_{\rm{RMS}}\ $}

\def\cm-3{\,{\rm cm^{-3}}}

\def\kpc-3{\,{\rm kpc^{-3}}}
\def\myr-1{\,{\rm Myr^{-1}}}

\def\kpc{\,{\rm kpc}}

\def\tc{$t_{\rm{c,0}}$}

\def\rsn{$R_{\rm{fade}}$}
\def\tmulti{$t_{\rm{multi}}$}

\def\vrms{$v_{\rm{RMS}}$}

\def\tcas{$t_{\rm{cas}}$}

\usepackage{natbib}
\usepackage{color}
\usepackage{rotating}
\usepackage{footnote}
\usepackage{datetime}
\usepackage{amsmath}
\usepackage{mathrsfs}
\usepackage[normalem]{ulem}
\shorttitle{SNe Ia Feedback \& ISM Turbulence}
\shortauthors{Li et al.}

\graphicspath{{./}{figures/}}

\begin{document}
\begin{CJK*}{UTF8}{gbsn}

\title{The Impact of Type Ia Supernovae in Quiescent Galaxies: \\ II. Energetics and Turbulence}

\correspondingauthor{Miao Li}
\email{mli@flatironinstitute.org}

\author[0000-0003-0773-582X]{Miao Li (李邈)}
\affiliation{Center for Computational Astrophysics, Flatiron Institute, New York, NY 10010, USA}

\author[0000-0001-5262-6150]{Yuan Li (黎原)}
\affiliation{Center for Computational Astrophysics, Flatiron Institute, New York, NY 10010, USA}
\affiliation{Department of Astronomy, and Theoretical Astrophysics Center, University of California, Berkeley, CA 94720, USA}

\author[0000-0003-2630-9228]{Greg L. Bryan}
\affiliation{Center for Computational Astrophysics, Flatiron Institute, New York, NY 10010, USA}
\affiliation{Department of Astronomy, Columbia University, 550 West 120th Street, New York, NY 10027, USA}

\author[0000-0002-0509-9113]{Eve C. Ostriker}
\affiliation{Department of Astrophysical Sciences, Princeton University, Princeton, NJ 08544, USA}

\author[0000-0001-9185-5044]{Eliot Quataert} 
\affiliation{Department of Astronomy, and Theoretical Astrophysics Center, University of California, Berkeley, CA 94720, USA}

\begin{abstract}
{
Type Ia supernovae (SNe Ia) provide unique and important feedback in quiescent galaxies, but their impact has been underappreciated. In this paper, we analyze a series of high-resolution simulations to examine the energetics and turbulence of the medium under SNe Ia. We find that when SN remnants are resolved, their effects differ distinctly from a volumetric heating term, as is commonly assumed in unresolved simulations. First, the net heating is significantly higher than expected, by 30$\pm$10\% per cooling time. This is because a large fraction of the medium is pushed into lower densities which cool inefficiently. Second, the medium is turbulent; the root-mean-squared (RMS) velocity of the gas to 20-50 km s$^{-1}$ on a driving scale of tens of parsec. The velocity field of the medium is dominated by compressional modes, which are larger than the solenoidal components by a factor of 3-7. Third, the hot gas has a very broad density distribution. The ratio between the density fluctuations and the RMS Mach number, parameterized as $b$, is 2-20. This is in contrast to previous simulations of turbulent media, which have found $b\lesssim$ 1. The reason for the difference is mainly caused by the \textit{localized} heating of SNe Ia, which creates a large density contrast. Last, the typical length scale of a density fluctuation grows with time, forming increasingly larger bubbles and filamentary ridges. These underlying density fluctuations need to be included when X-ray observations are interpreted.
 }

\end{abstract}

\keywords{Interstellar medium, Type Ia supernovae, Elliptical galaxies, Galaxy formation, Galaxy evolution, Hydrodynamical simulations, Shocks, Hot ionized medium}

\section{Introduction}
\label{intro}

Current quiescent galaxies formed most of their stars at $z\gtrsim$ 2. Since then, gas within galaxies has been mainly hot. The cooling time of this hot gas is short compared to the Hubble time, but massive ``cooling flows'' have not been detected \citep[see][for a review]{mathews03}. Thus, some feedback process must have been preventing gas from cooling.

Supermassive black holes (SMBHs) and SNe Ia dominate the energy input in these systems. Though SMBHs have been thought to play a major role, there is reason to believe that SNe Ia are important, even dominant, in some parts of the galaxies. SNe Ia occur throughout the galaxy, and their rate scales with the stellar density. While SMBH may dominate energy inputs at the center of galaxies and in limited conical regions surrounding jets if they are present, SNe Ia deposit more energy {\it locally} for most of the volume at larger radii. Jets from SMBH can reach large distances, but most energy is transferred along the jet, and it is unclear how the energy of the jet is converted to heat in the gas \citep{mcnamara07}. SMBHs are highly variable in their feedback power. In fact, most of the time they are dormant, and output energy much below the Eddington rate \citep{soltan82,kormendy13}. In contrast, the occurrence rate of SNe Ia is more steady, although it declines slowly over cosmic time \citep{cappellaro99,pain02,scannapieco05,maoz17}. Thus SNe Ia may be more important when the SMBH is less active. In addition, the output power of SMBH depends on their accretion rate, which is determined by the ISM conditions within the Bondi radius\citep{yuan14}. SNe Ia, by regulating the ISM around the SMBH, can play a significant role in determining the strength of SMBH feedback. Quantitatively, the SNe Ia heating rate roughly balances the radiative cooling rate of hot gas over a wide range of radii \citep[][and references therein]{voit15}.

Compared to SMBH feedback, which has been studied extensively in recent years \citep[e.g.][]{ciotti07,choi12,gaspari12,dubois13,yli15,yuan18,dave19}, SNe Ia feedback has been relatively neglected \citep[though see works by][]{tang05,tang09}. One of the reasons is that the resolution of cosmological simulations, or even isolated galaxy models, is insufficient to resolve individual SN remnants. Instead, SNe feedback is added as a subgrid heating term. The accuracy of this modeling has not been examined.  In addition, SNe also drive turbulence in the gas. This has important implications on how the heat and metals are transported \citep{mathews90,renzini93}. However, turbulence in the hot ISM of elliptical galaxies, driven by many discrete SNe explosions from small scales, has been studied little \citep{moss96}. On the other hand, future X-ray missions, such as Athena, Lynx, and HUBS, will provide unprecedented details on the dynamics and thermodynamics of the hot ISM and the circumgalactic medium. A thorough understanding of the physical processes in the hot gas is mandatory for predicting and interpreting observations.

This is the second in a series of two papers, in which we study SNe Ia feedback in the hot ISM typical of quiescent galaxies. We model a patch of the hot ISM and resolve individual SN remnants, to examine the effect of SNe Ia on the hot ISM. In particular, we evaluate what has been missing in coarse-resolution simulations where SNe Ia are underresolved. In \citet[][hereafter Paper I]{li20}, we focused on the formation of the cool gas in these systems. The randomly located SNe Ia heat the gas unevenly, and gas not covered by any SNe cools down. This can occur even when the overall heating rate of SNe is higher than the radiative cooling rate. 
In the present paper, we investigate how SNe Ia affect the thermodynamic evolution and the turbulence structure of the gas, and discuss its implications on X-ray observations.

We organize our paper as follows. In Section 2, we briefly recapitulate the numerical method. In Section 3, we examine the energy evolution and compare it to the subgrid modeling of SNe Ia in cosmological simulations. In Section 4, we investigate the turbulent structure of the hot gas. We summarize our results in Section 5 and provide the concluding remarks in Section 6. 

This is a paper from the \textit{Simulating Multiscale Astrophysics to Understand Galaxies} (SMAUG) collaboration\footnote{\url{www.simonsfoundation.org/flatiron/center-for-computational-astrophysics/galaxy-formation/smaug}}, a project intended to improve models of galaxy formation and large-scale structure by working to understand the small-scale physical processes that cannot yet be directly modeled in cosmological simulations.

\section{Methods}
\label{sec:method}

We have run a series of idealized simulations. Each simulation is carried out in a 3D box with dimensions of order 1 kpc, representing a small patch of an elliptical galaxy. For a detailed description of the simulation setup, see Section 2 of Paper I. 

The initial condition is a uniform and static medium, with a number density $n$ and temperature $T$. The range of densities and temperatures is based on the observed values in elliptical galaxies of different masses, and also from different radial locations inside elliptical galaxies \citep{voit15}. Periodic boundary conditions are applied. SNe are added at fixed intervals but at random locations in the box. Each SN has $E_{\rm{SN}}=10^{51}$ erg and 1 $M_\odot$, which are injected within a sphere of radius $0.5R_{\rm{fade}}$ ($R_{\rm{fade}}$ is the radius of the SN bubble when reaching pressure equilibrium with the ambient medium, see Eq. 2 of Paper I). For each simulation, the overall heating rate by SNe is $H\equiv S E_{\rm{SN}}$, where $S$ is the rate of SNe per time per volume. Observations indicate that the cooling rate of the hot medium is roughly balanced by the heating rate of SNe Ia \citep{voit15}, though the exact ratio is not known to within a factor of 2 due to uncertainties in the SNe Ia rate and gas metallicity. Therefore we vary $H/C$ around unity by about a factor of 2. Optically-thin radiative cooling is implemented for the temperature range of 300-10$^9$ K , with the cooling function $\Lambda (T)$ shown in Fig. A1 of Paper I. We define the cooling rate per volume for the initial condition as $C \equiv n^2\Lambda(T)$. 

We vary $n$, $T$ and $H/C$ for different runs, and each run is represented by a name showing its input parameters. For example,  ``n0.02-T3e6-H1.4C'' indicates $n=0.02$ cm$^{-3}$, $T=3\times 10^6$ K and $H/C=1.4$. For the names omitting $H/C$, the fiducial value  $H/C=$1.02 is used.  Table 1 (an extended version of Table 1 in Paper I) lists the parameters for all the runs. We emphasize that while the volumetric energy input rate $H$ and the mean density $n$ remain at their initial values, the temperature and density both vary throughout the box, and the cooling rate therefore is not constrained to remain at the initial value $C$. We ran the simulations for 4 cooling times for each set of initial conditions. The time at which the cool phase forms is indicated by \tmulti\ (listed in Table I), which is defined as the time when the first cell of the box cools below $2\times10^4$ K.

\linespread{0.9}
\begin{table*}[]
\begin{center}
\caption{Model Parameters}
\label{table1}
\begin{tabular}{cccccccccccc}
\hline
Name\footnote{Meaning of the symbols. S: volumetric  SNe Ia rate; d: galactocentric distance; $R_{\rm{fade}}$: fade-away radius for SN remnant;  $t_{\rm{c,0}}$: instantaneous cooling time; $t_{\rm{c,1}}$: integrated cooling time; $t_{\rm{multi}}$: formation time of cool phase;  $t_d$: turbulence decay time; $v_{\rm{RMS}}$: root-mean-squared velocity; $v_{\rm{comp}}$: compressional velocity;  $v_{\rm{sol}}$: solenoidal velocity; $\mathcal{M}_{\rm{RMS}}$: root-mean-squared Mach number. For detailed descriptions of these quantities, see \citep[][Paper I]{li20}}           & S                       & d     & $R_{\rm{fade}}$ & $t_{\rm{c,0}}$ & $t_{\rm{c,1}}$ & $t_{\rm{multi}}$ & $t_d$ & $v_{\rm{RMS}}$ & $v_{\rm{comp}}$ & $v_{\rm{sol}}$ & $\mathcal{M}_{\rm{RMS}}$ \\
                 & (Mpc$^{-1}$ kpc$^{-3}$) & (kpc) & (pc)            & (Myr)          & (Myr)          & (Myr)            & (Myr) & (km s$^{-1}$)         & (km s$^{-1}$)          & (km s$^{-1}$)         &                          \\\hline
n0.32-T1e7-H0.8C & 776                     & 0.3   & 19.4            & 33             & 28             & 72               & 7.0   & 28$\pm$15      & 28$\pm$14       & 7.1$\pm$7.0    & 0.06                     \\
n0.32-T1e7       & 990                     & 0.3   & 19.4            & 33             & 28             & 85               & 6.0   & 31$\pm$15      & 31$\pm$15       & 6.6$\pm$5.9    & 0.052                    \\
n0.32-T1e7-H1.2C & 1165                    & 0.3   & 19.4            & 33             & 28             & 85               & 5.3   & 31$\pm$17      & 31$\pm$16       & 6.7$\pm$6.2    & 0.05                     \\
n0.32-T1e7-H1.4C & 1359                    & 0.3   & 19.4            & 33             & 28             & --               & 4.7   & 33$\pm$16      & 33$\pm$16       & 7.8$\pm$7.1    & 0.046                     \\\hline
n0.16-T1e7       & 248                     & 0.5   & 24.5            & 67             & 57             & 180              & 11.0  & 28$\pm$14      & 28$\pm$14       & 5.5$\pm$5.2    & 0.046                    \\
n0.08-T1e7       & 61.9                    & 1     & 30.8            & 133            & 114            & 375              & 19.0  & 24$\pm$12      & 24$\pm$12       & 4.5$\pm$4.1    & 0.046                    \\
n0.02-T1e7       & 3.87                    & 4    & 48.8            & 534            & 456            & 1700             & 61.0  & 21$\pm$10      & 22$\pm$10       & 3.2$\pm$2.8    & 0.032                    \\\hline
n0.08-T3e6       & 110                     & 1     & 46.0            & 22             & 9              & 18               & 6.0   & 31$\pm$18      & 30$\pm$15       & 9.9$\pm$9.8    & 0.12                     \\
n0.08-T6e6       & 51.6                    & 1     & 36.5            & 96             & 56             & 180              & 16.5  & 24$\pm$13      & 24$\pm$12       & 5.1$\pm$4.9    & 0.055                    \\
n0.08-T3e7       & 92.3                    & 1     & 21.3            & 265            & 329            & 1162             & 25.0  & 30$\pm$13      & 30$\pm$14       & 3.9$\pm$3.4    & 0.03                     \\\hline
n0.02-T3e6       & 6.85                    & 4    & 72.9            & 90             & 35             & 79               & 20.0  & 23$\pm$13      & 23$\pm$12       & 6.1$\pm$5.8    & 0.087                    \\
n0.02-T3e6-hr    & 6.85                    & 4    & 72.9            & 90             & 35             & 82               & 28.0  & 25$\pm$14      & 25$\pm$13       & 7.2$\pm$6.9    & 0.092                    \\
n0.02-T3e6-H1.1C & 7.39                    & 4    & 72.9            & 90             & 35             & 84               & 19.0  & 25$\pm$13      & 25$\pm$12       & 6.5$\pm$6.1    & 0.085                    \\
n0.02-T3e6-H1.2C & 8.06                    & 4    & 72.9            & 90             & 35             & 95               & 18.0  & 25$\pm$13      & 25$\pm$12       & 6.2$\pm$5.8    & 0.082                    \\
n0.02-T3e6-H1.4C & 9.40                    & 4    & 72.9            & 90             & 35             & 155              & 16.0  & 25$\pm$13      & 25$\pm$12       & 6.3$\pm$6.1    & 0.074                    \\
n0.02-T3e6-H1.8C & 12.1                    & 4    & 72.9            & 90             & 35             & --               & 13.0  & 26$\pm$13      & 26$\pm$13       & 6.3$\pm$6.0    & 0.05                     \\\hline
n0.002-T3e6      & 0.0685                 & 20    & 157.2           & 905            & 351            & 850              & 130.0 & 16$\pm$8       & 16$\pm$8        & 2.7$\pm$2.5    & 0.046         \\\hline
\end{tabular}
\end{center}
\end{table*}

\section{Energy Evolution}

\subsection{Results}

We consider one computational cell in an unresolved simulation, where SN feedback is modeled as a subgrid heating source. The energy change at each time step is $\Delta E' = (H-C)\Delta t$. In other words, it assumes that the energy $E'$ evolves linearly with time,
\begin{equation}
    E'(t) = E_0\ \bigg[1+ \Big( \frac{H}{C}-1 \Big)\ \frac{t}{t_{\rm{c,0}} }  \bigg],
    \label{eq:E'}
\end{equation}
where \tc$=P/(\gamma -1)/C$.
In each of our simulation boxes, the internal structure is resolved by many individual cells, such that the density and temperature vary throughout the domain.
We measure the actual energy of the box, $E(t)$, and compare it with the linear model indicated in Eq. \ref{eq:E'}. 

\begin{figure}
\begin{center}
\includegraphics[width=0.50\textwidth]{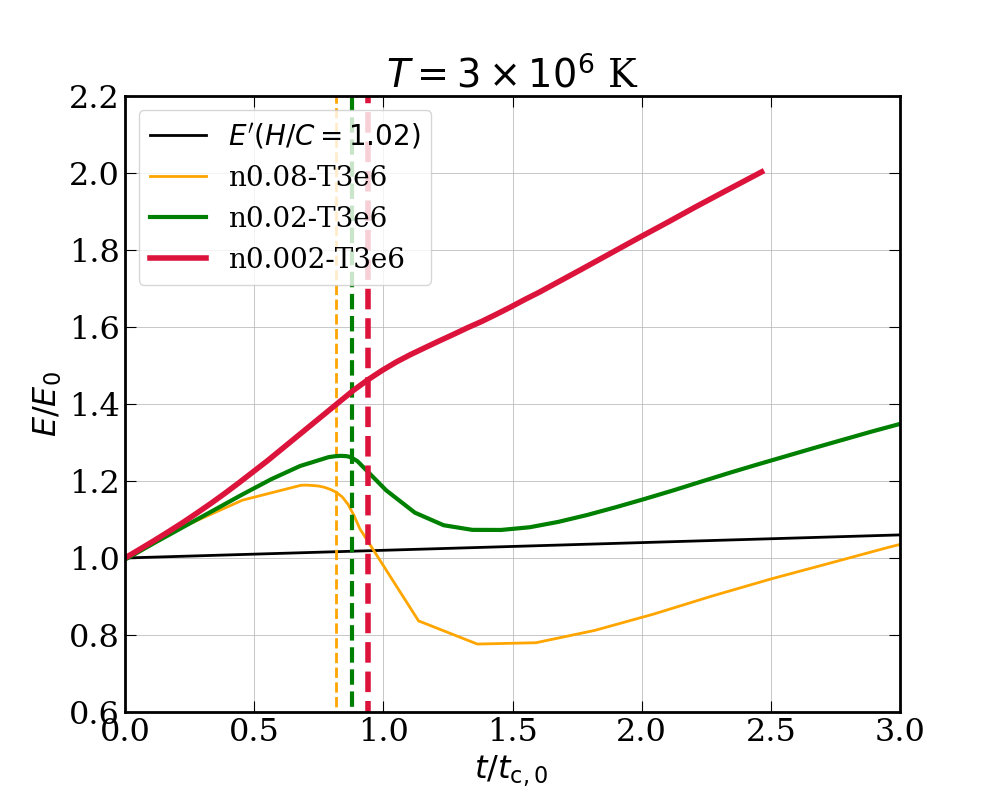}
\includegraphics[width=0.50\textwidth]{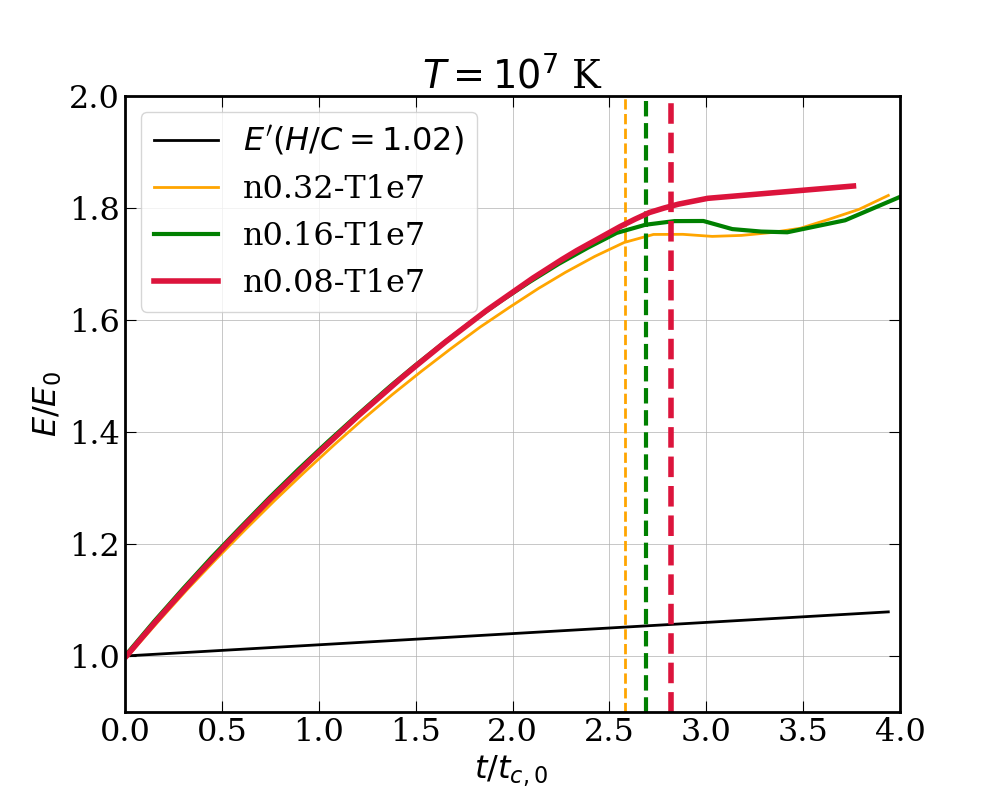}
\caption{Evolution of the total energy in the simulation box, normalized by the initial energy. Upper panel: runs with $T=3\times 10^6$ K. Lower panel: $T=10^7$ K. Different curves have different $n$. All of the runs have $H/C=1.02$. The vertical lines indicate \tmulti, the time when the cool phase forms. The solid black lines indicate $E'$ from Eq. \ref{eq:E'}, i.e., a linear change in energy with time for a fixed $H/C$. All simulations deviate from the simple linear evolution of energy.}
\label{f:e_t}
\end{center}
\end{figure}

Fig. \ref{f:e_t} shows the total energy $E$ in the box as a function of time. The energy is normalized by the total energy at $t=$0, $E_{\rm{0}}$. The black line indicates $E'$ in Eq. \ref{eq:E'}. The upper panel is for the runs with $T=3\times 10^6$ K and the lower for $T=10^7$ K. The runs have different $n$ but they all have the same (initial) $H/C=1.02$. The vertical dashed line indicates \tmulti, i.e., when the cool phase forms. The actual energy evolution deviates significantly from the linear function of time described by Eq. \ref{eq:E'}.  For cases with $T=3\times 10^6$ K, the energy evolution for different $n$ varies greatly, whereas for $T=10^7$ K, the trajectories are almost the same. Now we consider them in some detail.

For n0.02-T3e6, $E$ first increases, at a rate much higher than $H/C=$1.02. Immediately before the formation of cool phase, the total energy is 27\% higher than the initial value. Since the energy injection rate is constant, the rising of the energy implies that the actual radiative cooling is significantly suppressed compared to the initial condition. The energy then quickly declines when the cool phase forms. At about 1.4\tc, the energy reaches the lowest point (but still above $E'$), after which the energy increases again. 

For n0.08-T3e6, the shape of the curve $E(t)$ is quite similar to that of n0.02-T3e6, but the total energy is consistently lower (except for the very early stage at $t\lesssim 0.3$\tc). At about 0.9\tc, $E$ decreases below $E'$. In fact, the system has even less energy than the initial value, meaning that the energy loss from radiative cooling exceeds the energy injected by SNe. Then the energy starts to grow again, and by about \tc, the actual energy is close to $E'$. 

For n0.002-T3e6, there is no decrease of energy over time. Even after the cool phase forms, the energy still increases, although the rate of increase slows down after \tmulti. The total energy is much higher than $E'$: at \tc, $E \approx 1.5 E'$, and at 2\tc, $E \approx 1.8 E'$. The overall cooling is greatly suppressed.  

Interestingly, for cases with $T=10^7$ K, the energy evolution is almost independent of $n$, in contrast to the $3\times 10^6$K case. The total energy almost always increases, even after \tmulti\ (except for a brief period at about 3.2-3.6 \tc\ for n0.16-T1e7). By \tc, $E$ is  35\% more than $E'$; by 2\tc, $E$ is about 65\% more than $E'$. That said, the energy accumulation rate is greatly suppressed at $t\geqslant$\tmulti. 

\begin{figure}
\begin{center}
\includegraphics[width=0.5\textwidth]{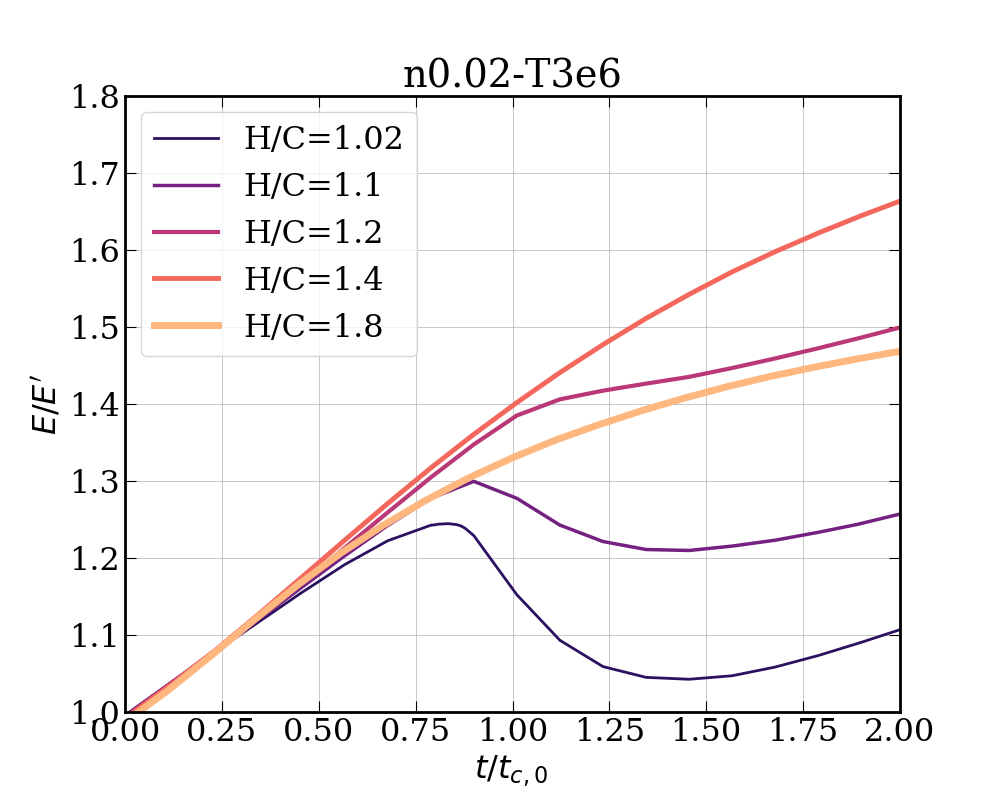}
\includegraphics[width=0.5\textwidth]{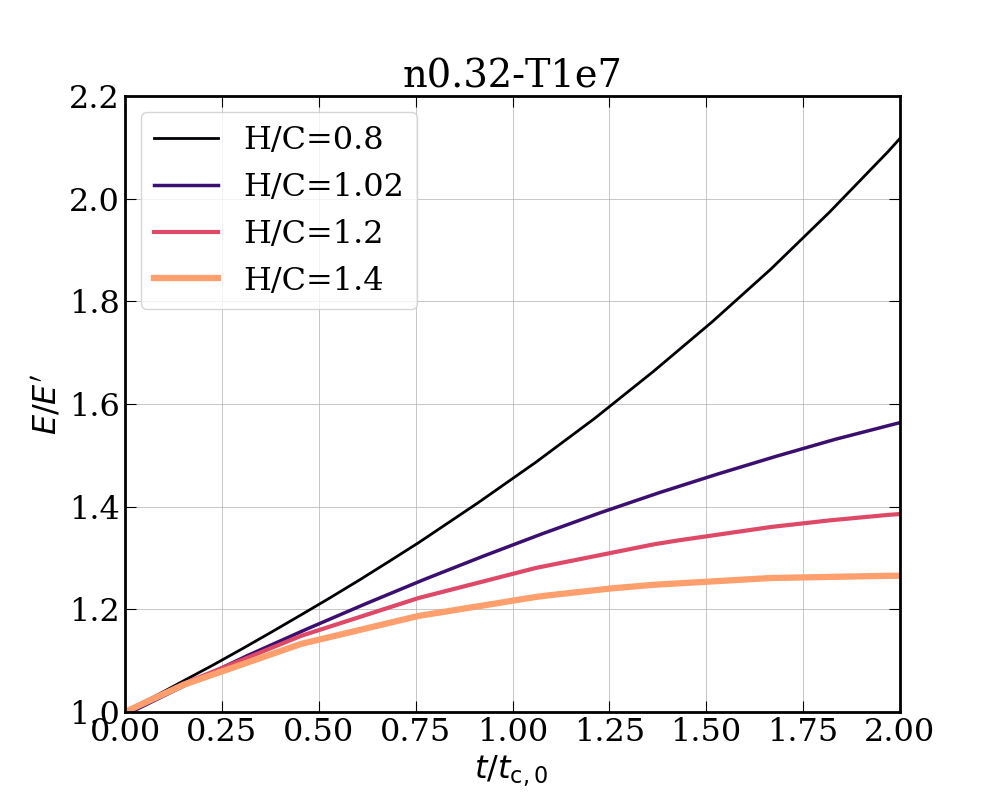}
\caption{Ratio between total energy in simulation box and $E'$ (Eq. \ref{eq:E'}) as a function of time for different $H/C$. Upper panel: n0.02-T3e6; lower panel: n0.32-T1e7. The actual energy evolution is consistently larger than $E' (t)$. 
}
\label{f:e_t_HC}
\end{center}
\end{figure}

Fig. \ref{f:e_t_HC} shows the change in energy for the same $n$ and $T$, but different $H/C$. Note that in contrast to Fig. \ref{f:e_t}, the y-axes in this figure show $E/E'$. The upper panel shows the runs with n0.02-T3e6, which have $H/C=$1.02, 1.1, 1.2, 1.4, and 1.8, respectively. We find that similar to the fiducial H/C$=$1.02, $E/E'$ is always above unity. The ratio  $E/E'$ is not a monotonic function of $H/C$. For $H/C<$ 1.8,  $E/E'$ is larger as $H/C$ increases. For $H/C=$1.8, the curve is actually lower than that of H/C$=$1.4 and 1.2. Generally, as in the cases shown in Fig. \ref{f:e_t}, the rate of energy increase slows down after \tmulti. 

The lower panel of Fig. \ref{f:e_t_HC} shows the cases with n0.32-T1e7,with $H/C=$0.8, 1.02, 1.2, 1.4, respectively. Similar to n.02-T1e6, all the actual energy evolution is above $E'$. Interestingly, the lower $H/C$ is, the higher $E/E'$ is, in contrast to the general trend for n0.02-T3e6. Overall, by \tc, the actual energy is larger than $E'$ by 10-50\%. This evolution of the energy is missed when SNe are treated as a subgrid heating term with low resolution.

In summary, Fig. \ref{f:e_t} and \ref{f:e_t_HC} show the diversity of the energy evolution of a hot patch of ISM when SNe are resolved. They differ greatly from the linear function of time indicated by Eq. \ref{eq:E'}. Except when $n$ is high and $T$ is low (like the case n0.08-T3e6), the total energy is above that of Eq. \ref{eq:E'}.

\subsection{Causes, Caveats and Implications}

\begin{figure}
\begin{center}
\includegraphics[width=0.50\textwidth]{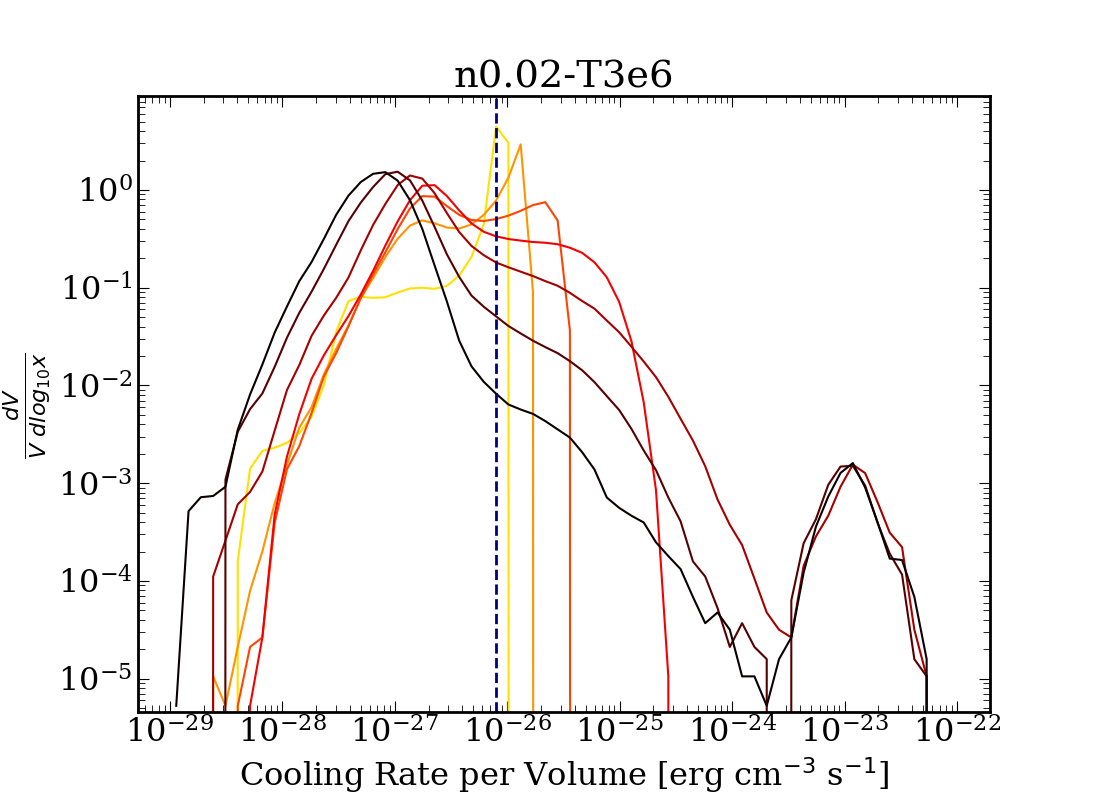}
\includegraphics[width=0.50\textwidth]{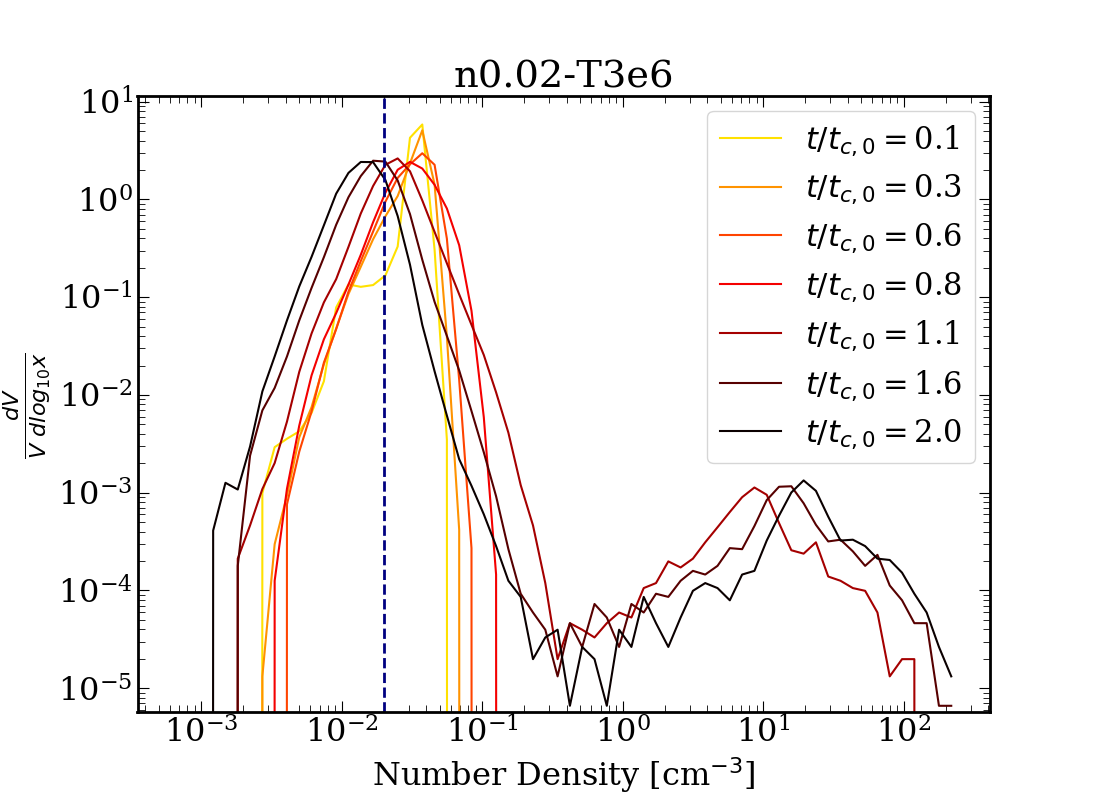}
\caption{Volume probability distribution function (vPDF) of logarithmic cooling rate per volume (upper panel) and logarithmic density (lower panel) at different times for n0.02-T3e6. The dashed lines indicate the initial condition. Both distributions broaden significantly over time.}
\label{f:cool_rate_den_dist}
\end{center}
\end{figure}

So far, we have seen the empirical energy evolution of the hot gas, where the radiative cooling is generally suppressed compared to the uniform condition. To better understand the energy evolution, we plot in Fig. \ref{f:cool_rate_den_dist} the volume probability distribution function (vPDF) of gas properties for n0.02-T3e6 at different times. The upper panel shows the PDF of logarithmic cooling rate per volume, $n^2 \Lambda (T)$, while the lower one shows that of logarithmic number density. The initial vPDFs are delta functions peaked at $8\times 10^{-27} $ erg cm$^{-3}$ s$^{-1}$ and 0.02 cm$^{-3}$, respectively, indicated by the dashed lines. Soon after the simulation begins, both vPDFs widen significantly. An increasingly larger volume is pushed to the regime where density and cooling rate are low. The shape of the two vPDFs for the same snapshot is similar, although the cooling rate has a broader distribution than the density. This results from the heating and rarefaction effect of SNe-driven blast waves. At the same time, some gas has an enhanced density and cooling rate compared to its initial condition. This is the gas that is not covered by any SNe bubbles. The dense and high-cooling rate wing of the vPDF eventually forms the cool phase. The broader vPDF of cooling rate can be understood as follows: (1) the cooling rate per volume has a factor of $n^2$, and (2) the cooling curve $\Lambda(T)$ has a negative slope at $T\approx 3\times10^6$ K, thus, assuming pressure equilibrium, the cooling rate is further reduced for hot/low-density gas. Another way of thinking is that in this temperature regime ($10^{5-7}$ K), the cooling time is roughly proportional to entropy $K^{3/2}$; and by generating entropy, SNe shocks decrease the net cooling rate. 

To summarize, we have found that for the hot ISM in elliptical galaxies, SNe Ia can lead to a very broad density distribution, and an even broader distribution of cooling rate. Overall, the cooling rate of the medium is more suppressed than that of a uniform medium. The difference is 10-50\% at \tc. The overheating holds for a broad set of conditions of density and temperature. This means that the overheating can occur for a significant fraction of the elliptical galaxies. This will lead to a global expansion of the hot ISM and generation of galactic outflows. The enriched outflows by SNe Ia can interact with the circumgalactic medium and leave chemical imprints there \citep{chen18,zahedy19}. The long-term evolution of overheated hot gas and its implications for the evolution of elliptical galaxies should be investigated in future studies. 

This overheating of the medium may cause concern regarding the adoption of a periodic boundary condition, which does not allow the gas to go beyond the boundaries. Indeed, in a realistic environment, an overpressured patch should expand, provided that the neighboring patches do not add pressure at the same rate. However, we emphasize that a criterion for the onset of overheating is independent of the boundary condition adopted, but is due to the medium and localized heating of SNe. Furthermore, during our simulation period, the overpressure is generally mild, so the confining effect by the periodic conditions is not very strong and can be quantified (as we will show in the sections below). The bottom line is that while a periodic boundary is clearly unsuitable for the long-term evolution of the hot ISM under thermal runaway, useful results can still be extracted before the system evolves too far from the initial condition.

We also note the difference between the overheating effect by SNe we see here and that for disk galaxies, where the ISM components are very different. In disk galaxies, the ISM is predominantly in cooler phases and the SNe rate is determined by the star formation rate. In numerical simulations with sufficiently high SN rate such that remnants overlap, a local thermal runaway of the ISM will occur \citep[e.g.][]{li15,gatto15}. 
However, the extent of the thermal runway is quite different for the two cases:  the overheating is much more prominent for a disk galaxy ISM. For example, the gas pressure rises quickly by a factor of 10 within a few tens of Myr for a solar neighborhood condition \citep[Fig. 10 of][]{li15}; in contrast, the overheating we see in the elliptical environment typically leads to an increase of a factor of two (or less) in the pressure over a few hundred million years. The difference is that the SN rate is usually far higher in disk galaxies, and the resultant heating rate is far larger than the cooling rate of the hot phase, whereas in elliptical galaxies, the SN Ia heating is roughly equal to the cooling rate of hot gas.

In cosmological simulations, this overheating effect is missing since SNe Ia are generally unresolved. Therefore, the simple prescription of SN Ia feedback indicated by Eq. \ref{eq:E'} is not accurate; using this formula, the actual effect of SNe Ia feedback is underestimated by 10-50\% for a timescale of \tc. This may mean that some other forms of feedback, such as AGN, have to be artificially enhanced to compensate for the cooling of the medium, in order to suppress the cooling flows. 

To include the unresolved inhomogeneity in cosmological simulations, one may use a modified cooling rate calibrated from resolved simulations. It is challenging to formulate a simple recipe due to the variety of the energy evolution under different conditions (see Fig. \ref{f:e_t}, \ref{f:e_t_HC}). But a multi-dimensional look-up table can be constructed for this purpose, which we postpone for future work. 

\section{Turbulence}

\subsection{Velocity Structure and Turbulent Cascade}
\label{sec:velocity}

In this section we discuss the kinematics of the hot medium under the influence of SNe Ia.

\begin{figure}
\begin{center}
\includegraphics[width=0.45\textwidth]{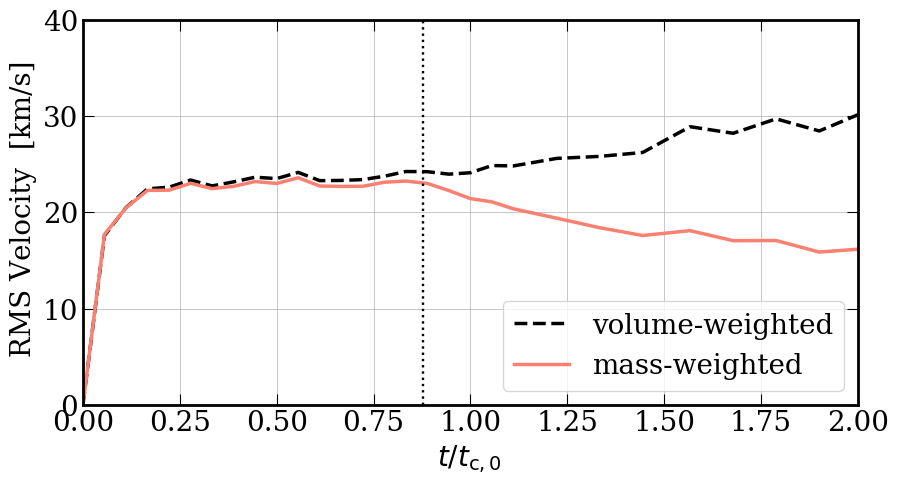}
\caption{Root-mean-squared velocity as a function of time for n0.02-T3e6. The solid line is weighted by volume and the dashed line is weighted by mass. The vertical dotted line indicates the time when the cool phase forms. }
\label{f:v_t}
\end{center}
\end{figure}

Fig. \ref{f:v_t} shows the RMS velocity \vrms\ as a function of time for n0.02-T3e6. The solid line denotes the volume-weighted velocity and the dashed line represents the mass-weighted velocity. The vertical dotted line indicates \tmulti. Shortly after the simulation starts, at about 0.2\tc\ (a few sound crossing times of the box), both velocities come to roughly constant values, which are very close to each other. Once the cool phase forms, the mass-weighted \vrms\ starts to decline while the volume-weighted one rises. This is because the volume-weighted \vrms\ reflects the velocity of the hot phase, whereas the mass-weighed \vrms\ skews to that of the cool phase. The cool phase, which has a small \vrms, occupies a tiny fraction of the volume but a significant fraction of the mass (for the mass fraction, see Fig. 4 of Paper I). As the mass accumulates into cool clumps, hot gas becomes more tenuous. Thus, it is increasingly easier for SNe to stir the hot gas, but harder to push the cool phase with its much higher inertia. Other simulation runs show a very similar velocity evolution.

\begin{figure}
\begin{center}
\includegraphics[width=0.5\textwidth]{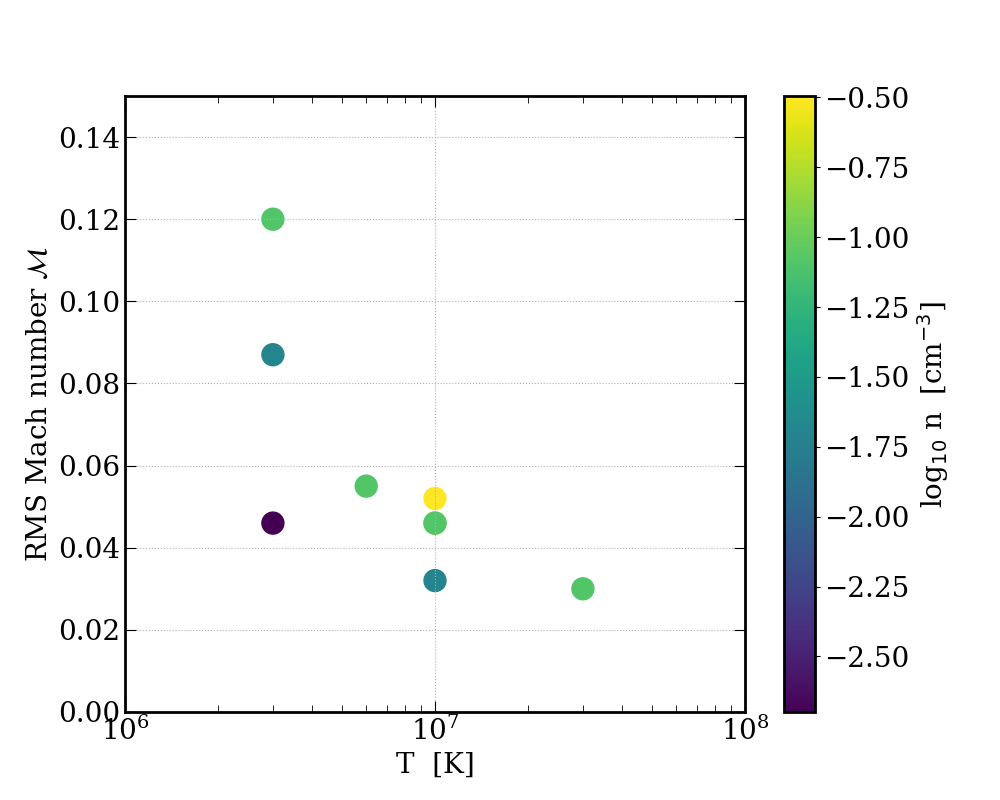}
\caption{Root-mean-squared Mach number of hot gas $\mathcal{M}$ prior to multiphase formation. The Mach number is about 0.1 for all cases.}
\label{f:Ia_all_v_RMS}
\end{center}
\end{figure}

Table 1 lists the volume-weighted \vrms\ for all the runs with $H/C=1.02$. The measurement is made prior to \tmulti\ but after the velocities reach the steady state. The mean values are 15-35 km s$^{-1}$ for all simulations. The standard deviation of the spatial fluctuation is about half the mean value. Compared to the adiabatic sound speeds of the medium, $c_s$, which are 200-500 km s$^{-1}$, \vrms\ is quite small. In Fig. \ref{f:Ia_all_v_RMS}, we show the RMS Mach number of the gas, with the Mach number $\mathcal{M}\equiv v$/$c_s$ for each cell. The Mach numbers are measured and averaged for the same duration as the \vrms. 
The Mach numbers are 0.03-0.13, which are quite small \footnote{Note that because of the overheating of the gas and the confinement by periodic boundaries, the sound speed increases over time. However, the change is moderate at most. For example, the pressure increases  by less than a factor of 2 (Fig. \ref{f:e_t}). Because the Mach number scales as $P^{-1/2}$, the change (decrease) in Mach numbers is small, less than 50\%. So the overall conclusion about the small Mach numbers does not change due to the application of periodic boundaries.}. Moreover, when we increase $H/C$ from 1.02 to 1.8 (for the case n0.02-3e6), the increase of \vrms\ is only a few km s$^{-1}$ (not shown in the figure, but see Table 1). This suggests that SNe can only drive very mild subsonic motions overall in the hot ISM of elliptical galaxies. This may seem counterintuitive since SNe explosions start with blast waves. However, the time for blast waves to decay into sound waves is very short, so the majority of the gas at a certain snapshot does not experience shocks and thus has small $\mathcal{M}$. The diagram also shows that conditions with lower $T$ and higher $n$ have a larger $\mathcal{M}$.

\begin{figure}
\begin{center}
\includegraphics[width=0.55\textwidth]{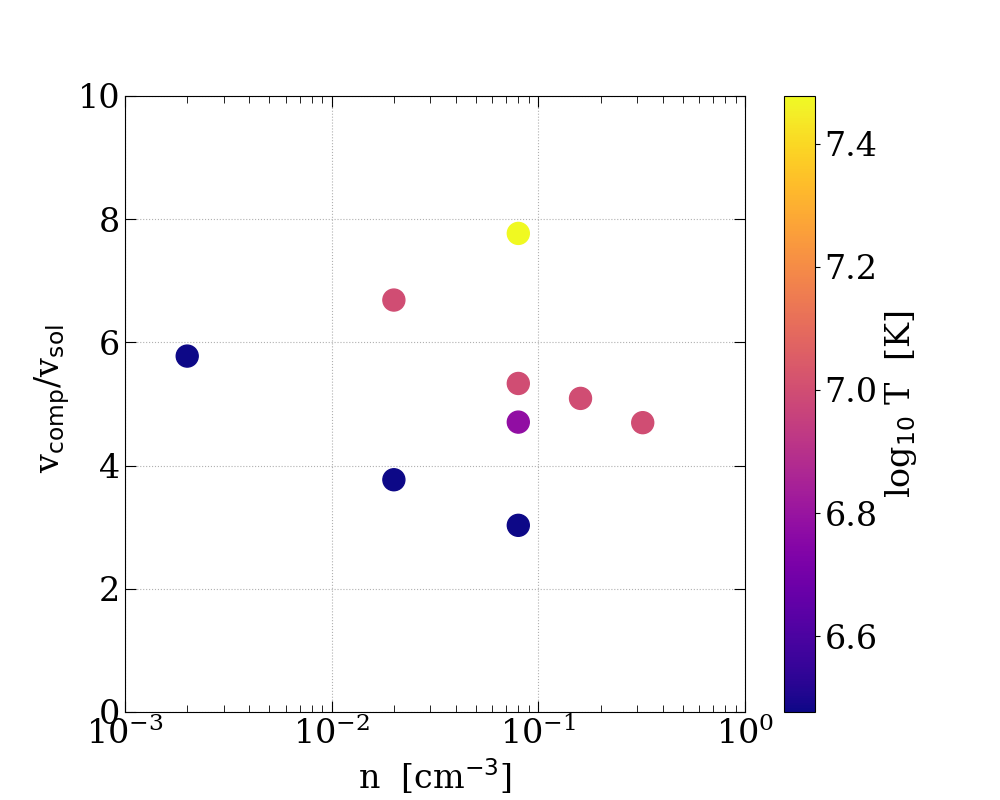}
\caption{Ratio of the compressional velocity component to the solenoidal velocity one, $v_{\rm{comp}}/v_{\rm{sol}}$, for all simulations with $H/C=$1.02. The ratio is 3-8, indicating that the compressional mode dominates over the solenoidal one.}
\label{f:Ia_all_v_com_v_sol}
\end{center}
\end{figure}

To better understand the nature of the motions driven by SNe Ia, we decompose the gas velocity into a divergence-free field and a curl-free one. Table 1 lists the magnitude of the compressional (curl-free) component, $v_{\rm{comp}}$, and that of the solenoidal (divergence-free), $v_{\rm{sol}}$, for all runs. Fig. \ref{f:Ia_all_v_com_v_sol} shows the ratio $v_{\rm{comp}}/v_{\rm{sol}}$, for simulations with $H/C=$ 1.02. The ratio is averaged over the same duration as \vrms, and we have found little time evolution. The ratio is 3-8, indicating that the compressional mode dominates over the solenoidal one. This is not surprising given that SNe initially only drive spherical, outward motions, which have zero vorticity. The solenoidal component arises because motions induced by different SNe interact with each other. Therefore, the compressional velocity is a first-order effect while the solenoidal is a second-order one. 

\cite{moss96} argued that vortical motions of hot gas are negligible under the impact of SNe Ia, because according to Kelvin's theorem, vortices cannot be created in inviscid flows. However, we do find considerable vortical motions, with an amplitude of 20\% \vrms. This difference is possibly because Kelvin's theorem only applies to barotropic fluids, which does not describe the present case as gas experiences radiative cooling and SNe heating. Moreover, the gas in the simulations is not entirely inviscid due to numerical viscosity.

It is found that in the ISM of disk galaxies, turbulence incurred by SNe is dominated by solenoidal motions \citep[e.g.][]{korpi99,balsara04,kapyla18}. This is in contrast with what we find with the hot medium in early-type galaxies. To convert the spherical blast waves into vortical motions, one needs certain conditions such as highly inhomogeneous medium, and/or sufficient interactions among multiple SN remnants. The ISM in elliptical galaxies is much smoother than that of disk galaxies, especially given that the cool gas only occupies a tiny volume, and the sparsity of the SNe explosions makes it hard for the remnants to overlap (see the discussion in Section 2 in paper I). Consequently, the velocity field in the hot ISM in these systems remains compression-dominated.

\begin{figure}
\begin{center}
\includegraphics[width=0.51\textwidth]{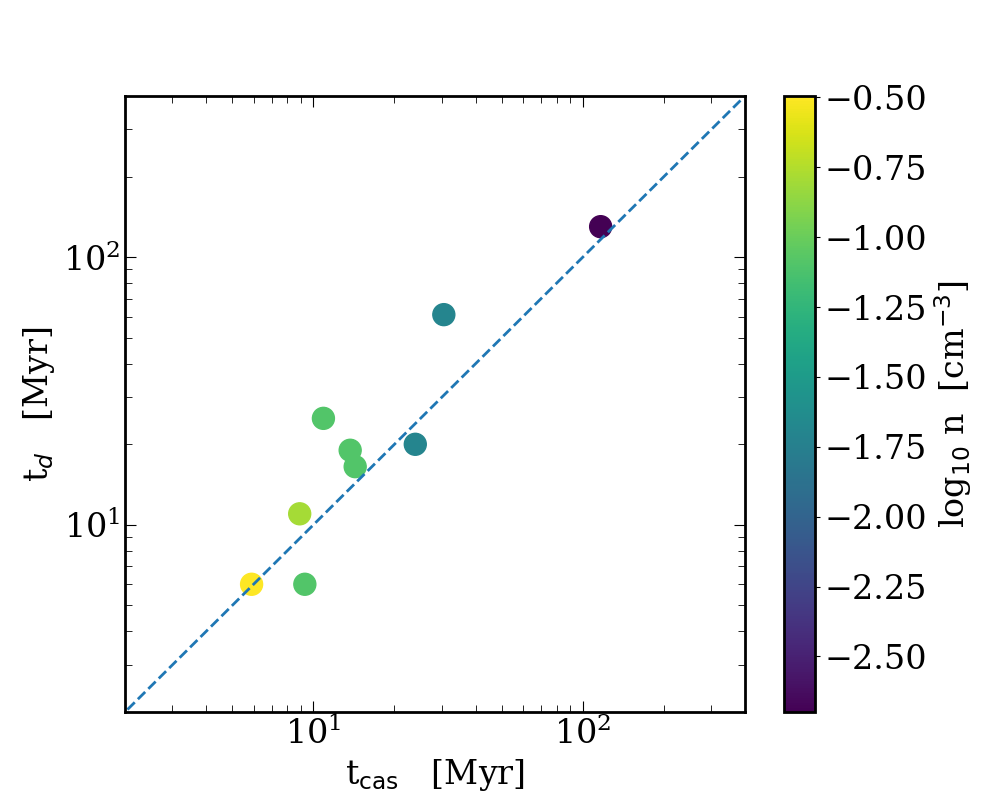}
\caption{Comparison of the turbulent cascade time, \tcas, with the empirically measured mixing time, $t_d$, for all simulations with $H/C=$1.02. The dashed line indicates \tcas$/t_d=$1. }
\label{f:t1_td}
\end{center}
\end{figure}

Given the measured solenoidal velocity, we can estimate the timescale for turbulent cascade, \tcas. The driving scale of the turbulence is  $\sim 2$\rsn. Therefore,  
\begin{equation}
   t_{\rm{cas}} \sim \frac{2R_{\rm{fade}}} {v_{\rm{sol}}}.
   \label{eq:tcas}
\end{equation}
This can be compared to the empirically measured mixing time, $t_d$, which is defined as the time it takes for the mass fraction of the unheated gas to become $e^{-1}$. The values of $t_d$ are listed in Table 1 (see also Fig. 9, 10 of Paper I). 
Fig. \ref{f:t1_td} shows the ratio \tcas\ versus $t_d$ for all simulations with $H/C=$ 1.02. The color of each circle shows $n$. The diagonal line indicates \tcas$=t_d$. All the data points are close to the diagonal line within a factor of a few, meaning that the empirical mixing time $t_d$ is well-described by the turbulent cascade time from Eq. \ref{eq:tcas}.

In Paper I, we found that turbulent mixing is more important in a hotter medium. That is, the ratio of the cooling time to the empirical mixing time, $t_{\rm{c,0}}/t_{d}$, is higher with higher gas temperature. Here we present an analytic formalism to illustrate this point. 
The system has two equilibria: \\
 (1) Balance of energy: 
 \begin{equation}
     H \sim C; \ \ \
\end{equation}
that is, 
\begin{equation}
     SE_{\rm{SN}} \sim n^2 \Lambda,   
     \label{eq:energy_bal}
 \end{equation}
 where $\Lambda$ is the cooling rate per particle.  \\
 (2) Balance of the injection and the dissipation of the turbulence, 
 \begin{equation}
 t_{\rm{inj}} \sim t_{\rm{cas}}\ , 
    \label{eq:turb_bal}
 \end{equation}
where the injection time scale
 \begin{equation}
     t_{\rm{inj}}= \frac{\rho v_{\rm{sol}}^2}{H f_{\rm{Ek}}} ,
 \end{equation}
 and $f_{\rm{Ek}}$ is the fraction of SN energy that contributes to the solenoidal motions. The dissipation time \tcas\ is from Eq. \ref{eq:tcas}, where a dependence $R_{\rm{fade}} \propto (nT)^{-1/3}$ is given in Eq. 3 of Paper I.
 
From Equations. \ref{eq:tcas}-\ref{eq:turb_bal}, and using  a power-law cooling function, $\Lambda \propto T^{-\alpha}$, we obtain
\begin{equation}
    \frac{t_{\rm{c,0}}}{t_{\rm{cas}}}  \propto n^{-\frac{4}{9}} T^{\beta} f_{\rm{Ek}}^{\frac{1}{3}}\ ; \ \ \ \beta = \frac{11}{9} + \frac{2\alpha}{3}\,
    \label{eq:tc_td}
 \end{equation}
For $10^5 \lesssim T \lesssim 10^7$ K, $\alpha\approx$ 0.7, therefore $\beta \approx$ 1.7. Eq. \ref{eq:tc_td} is consistent with the trend found in Paper I: $t_{\rm{c,0}}/t_{d}$ increases with decreasing density and increasing temperature, and it is much more sensitive to temperature than to density. Yet, the power-law indices of the fit to the simulation data, $t_{\rm{c,0}}/t_{d} \propto n^{-0.16\pm 0.02} T^{0.47\pm 0.03}$, are smaller than those of Eq. \ref{eq:tc_td}. This may be related to the unknown dependence of $f_{\rm{Ek}}$ on $n$ and $T$, and also to the relatively sparse population of the data on the $n-T$ plane). 

\subsubsection{Comparison to turbulence driven by other processes}

Turbulence is an important phenomenon in galaxies.
So far we have discussed the turbulence driven purely by SNe Ia. In a real environment, other processes can contribute to gas motions as well, such as AGN feedback, galaxy mergers, stellar winds, etc. One way to quantify their relative importance is to compare the turbulent velocity, or equivalently, the turbulent cascade time, at the same length scale.
Assuming (i) the driving scale is $l$, and the solenoidal velocity on that scale is $v_{\rm{sol,l}}$, (ii) turbulent cascade is fully established with a power-law spectrum for kinetic energy, i.e., $\widetilde{E}(k) \propto k^{-\beta}$, and (iii) the energy cascade rate for the inertial range is constant, then the turbulent cascade time for $l'<l$ is 
\begin{equation}
    \frac{ t_{\rm{cas}}'}{ t_{\rm{cas}}} = \left(\frac{l'}{l} \right)^{\beta-1}.
\end{equation}
 
Evaluating $t_{\rm{cas}}'$ for a Kolmogorov energy spectrum where $\beta = 5/3$, we obtain
\begin{equation}    
        t_{\rm{cas}}'= 8.2\ \rm{Myr} \left( \frac{l'}{100 \rm{pc}} \right)^{2/3} \left( \frac{l}{20 \rm{kpc}}\right)^{1/3} \left( \frac{70 km\ s^{-1}}{v_{\rm{sol,l}}} \right).
    \label{eq:tcas'}
\end{equation}
This can be compared to the turbulent cascade time driven by SNe Ia, $t_{\rm{cas}}$, which we find to be in the range of 5-120 Myr (see Table I). Observations have not been able to tightly constrain the turbulent velocities for giant elliptical galaxies \citep[see a recent attempt by][]{ogorzalek17}. Theoretically, for example, \cite{wang19} have found that AGN-driven turbulence on $l\approx$ 20 kpc has a velocity dispersion of 70 km s$^{-1}$. On scales of $l'\approx$ 2\rsn, $t_{\rm{cas}}'\approx$8 Myr, comparable to SNe Ia-driven turbulence. A similar magnitude of turbulence has been found in other AGN feedback models \citep{gaspari12,valentini15}. Note that since the solenoidal component is a fraction of the velocity dispersion, the estimated $t_{\rm{cas}}'$ is a lower limit.

Stellar winds from evolved stars can also contribute to the turbulence. This is due to the stellar velocity dispersion, which is typically several hundred km s$^{-1}$. \cite{mathews90} and \cite{moss96} estimated that the average turbulent velocity is about a few km s$^{-1}$ on a scale of one parsec. This is somewhat larger than what SNe Ia can drive on the same scale. However, as \cite{moss96} pointed out, the volume filling fraction of stellar winds is very small, about 10$^{-4}$, so the turbulence caused by stellar winds is highly inhomogeneous and not applicable to the general ISM. In contrast, turbulence driven by SNe Ia fills the volume and is more uniform spatially.

\subsection{Relation between density fluctuation and Mach number}

A simple correlation between density variation and the Mach number exists for a turbulent, isothermal medium,
\begin{equation}
\frac{\sigma_{\rho,V}}{\bar{\rho}_V} = b_\rho \mathcal{M}, 
\label{eq:rho_bM}
\end{equation}
where $\bar{\rho}_V$ is the volume-weighted mean density, and $\sigma_{\rho,V}$ is the volume-weighted standard deviation of density, $\mathcal{M}$ is the RMS Mach number, and $b_\rho$ is a proportional parameter \citep{padoan97}. When the medium has a log-normal density vPDF, which is empirically true for the isothermal turbulent medium, Eq. \ref{eq:rho_bM} is equivalent to
\begin{equation}
    \sigma_s^2 = \rm{ln}(1+b_s^2 \mathcal{M}^2),
    \label{eq:sigma_s}
\end{equation}
where $\sigma_s^2$ is the variance of the logarithmic density, and $b_\rho = b_s$ \citep[see][and references therein]{konstandin12}.

Numerical simulations of turbulent medium have found $b\lesssim$ 1. When gas is isothermal, the value of $b$ is closely related to the driving mechanism. When turbulence is driven purely by compressional mode, $b\approx$ 1; when driven solely by solenoidal mode, $b\approx 1/3$ for a supersonic medium, while $b$ becomes even smaller with decreasing $\mathcal{M}$ for a subsonic medium \citep{federrath08,federrath10,konstandin12,pan19}. 

Turbulence under non-isothermal conditions has also been investigated, adopting a polytropic equation of state \citep{li03,nolan15,federrath15}, or using a more realistic cooling/heating function \citep{wada01,audit10,gazol13}, or having additional physics such as magnetic fields, thermal conduction, gravitational stratification \citep{molina12,gaspari14}. Nevertheless, when using Eq. \ref{eq:rho_bM} or \ref{eq:sigma_s} to evaluate $b$, it is found that $b\lesssim$ 1 is still true. 

In this Section, we examine the correlation between $\sigma_{\rho,V}$ and $\mathcal{M}$ in our simulations, and compare to the previous work.

\begin{figure}
\begin{center}
\includegraphics[width=0.5\textwidth]{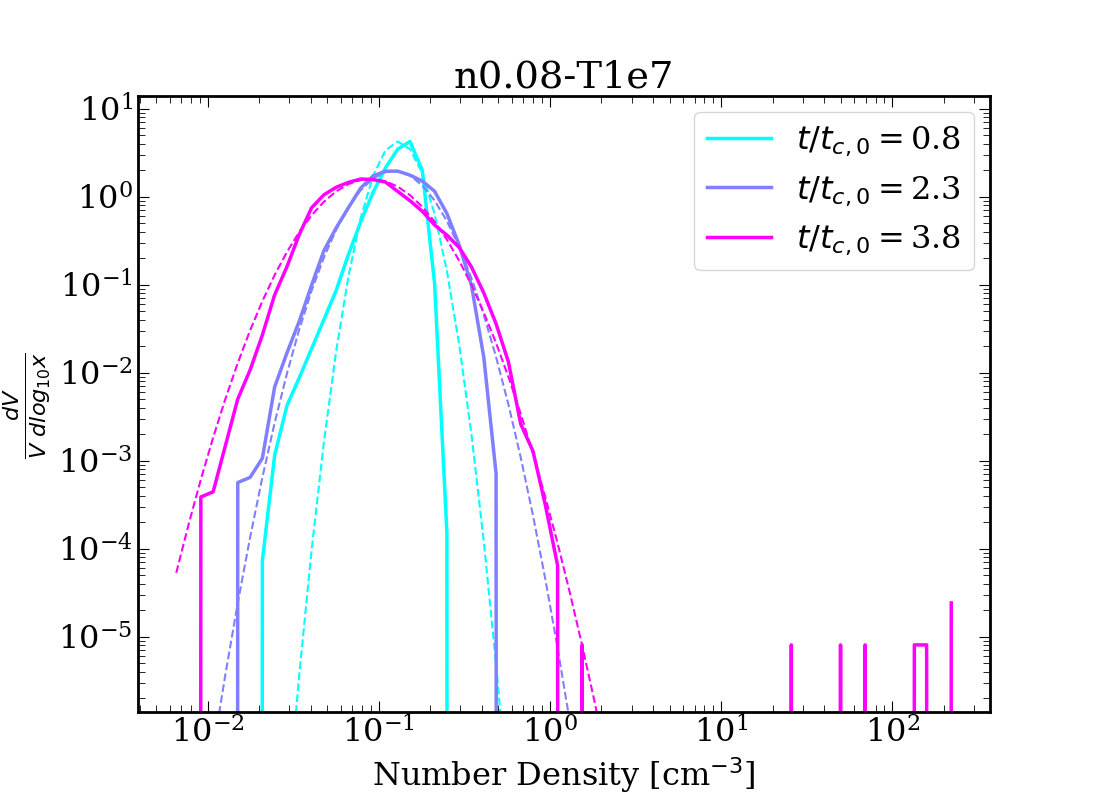}
\caption{Volume probability distribution function of logarithmic density for n0.08-T1e7. The solid lines are simulation outputs. The dashed lines of the same color indicate a log-normal function that has the same mean and standard deviation as those of the hot gas from the simulations. At late times, the hot gas density distribution is close to log-normal. }
\label{f:den_lognormal}
\end{center}
\end{figure}

First, we show that the distribution function of gas density does evolve toward a log-normal function. 
Fig. \ref{f:den_lognormal} shows the vPDFs of logarithmic density for a few snapshots of n0.08-1e7. The solid lines indicate the simulation outputs. The dashed lines with the same color show a log-normal distribution that has the same mean and standard deviation as the logarithmic density of the hot gas from the simulations. (For the last snapshot, cool gas has formed, but we do not include this phase when calculating the mean and standard deviation.) At early time, the density distribution is skewed and does not follow a log-normal shape. But as time goes by, the vPDF becomes more symmetric and approaches the log-normal distribution, while undergoing significant broadening. The evolution toward a log-normal distribution is seen in other simulations as well.

\begin{figure}
\begin{center}
\includegraphics[width=0.50\textwidth]{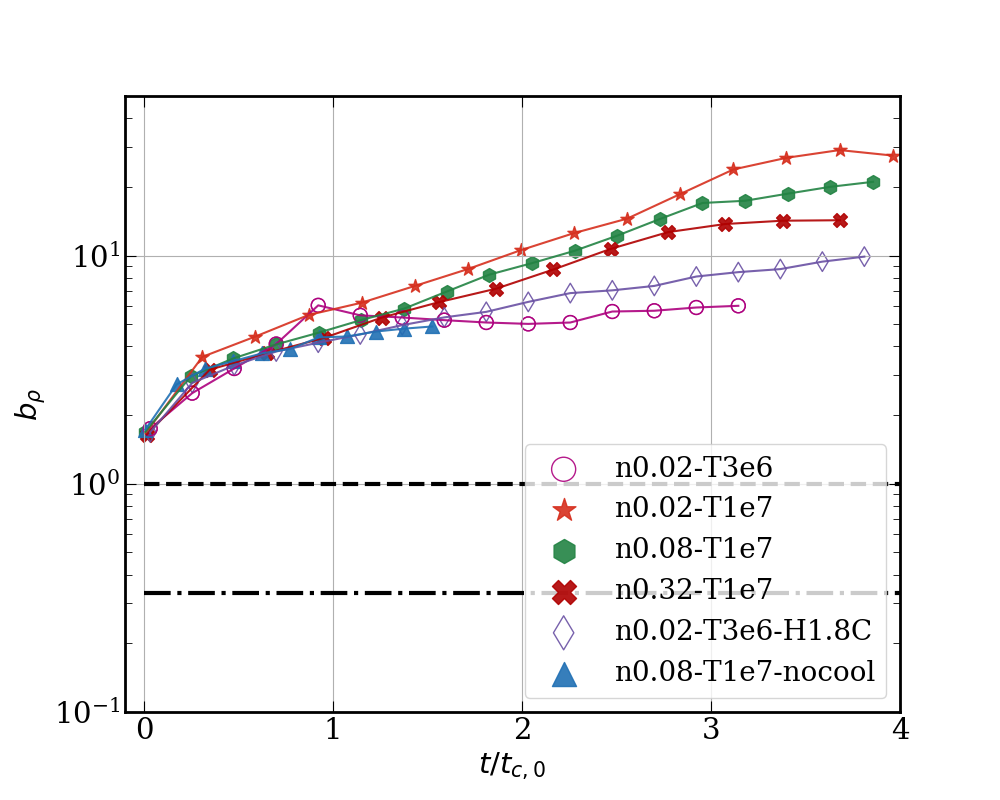}
\includegraphics[width=0.46\textwidth]{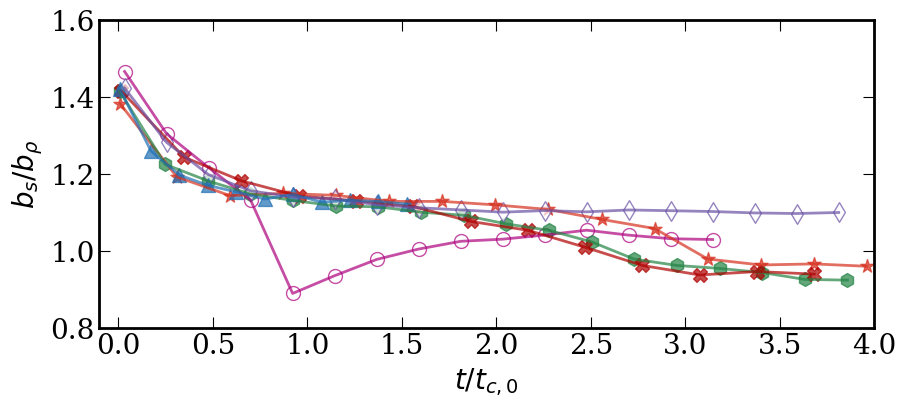}
\caption{Upper panel: Measured $b_\rho$ for hot gas as a function of time for a sample of runs. The values of $b$ are larger than 1 from early on, and increase with time. For supersonic isothermal gas, $b\approx 1/3$ is found for pure solenoidal driving (dot-dashed line) and $b\approx$1 for pure compressional driving (dashed line). Gas in our simulations has a significantly larger $b$ than those from an isothermal turbulent medium. Lower panel: ratio $b_s/b_\rho$ versus time for the same runs as in the upper panel. }
\label{f:b-t}
\end{center}
\end{figure}

Fig. \ref{f:b-t} shows the measured $b_\rho$, obtained from Eq. \ref{eq:rho_bM}, as a function of time for a sample of runs. The data are drawn from hot gas with $T>3\times 10^5$ K. The first data point for each run is taken at 1-5 sound crossing times of the simulation box, and the data sampling is uniform in time.
We find that even from very early on, $b_\rho \geqslant$ 1.5, and as time goes by, $b_\rho$ increases almost monotonically. The brief spike of n0.02-T3e6 at 0.9 \tc\ corresponds to the time when the cool phase forms, which causes a temporally broadened density range of the selected gas. 

At later times, runs with lower $n$ and higher $T$ have larger $b_\rho$ at a given $t/$\tc. This can be seen by comparing the three runs with $T=10^7$ K and $n=$ (0.32, 0.08, 0.02) cm$^{-3}$, and the two runs with $n=0.02$ cm$^{-3}$ and $T=$ ($3\times 10^6$, $10^7$) K. When cooling is turned off (n0.08-T1e7-nocool), $b$ is somewhat smaller than the fiducial run (n0.08-T1e7), but still much larger than 1.

The lower panel of Fig. \ref{f:b-t} shows the ratio $b_s/b_\rho$ as a function of time for the above runs, where $b_s$ is obtained from Eq. \ref{eq:sigma_s}. The ratio ranges from 0.9-1.5, which means that $b_s$ and $b_\rho$ are quite similar to each other, and generally, larger $b$ is found when the equation for the logarithmic density is used. The ratio is closer to unity at $t\gtrsim$\tc, when the density distribution is closer to log-normal. This indicates that $b \gg 1$  is robust regardless of whether Eq. \ref{eq:rho_bM} or \ref{eq:sigma_s} is used.

\begin{figure*}
\begin{center}
\includegraphics[scale=0.55]{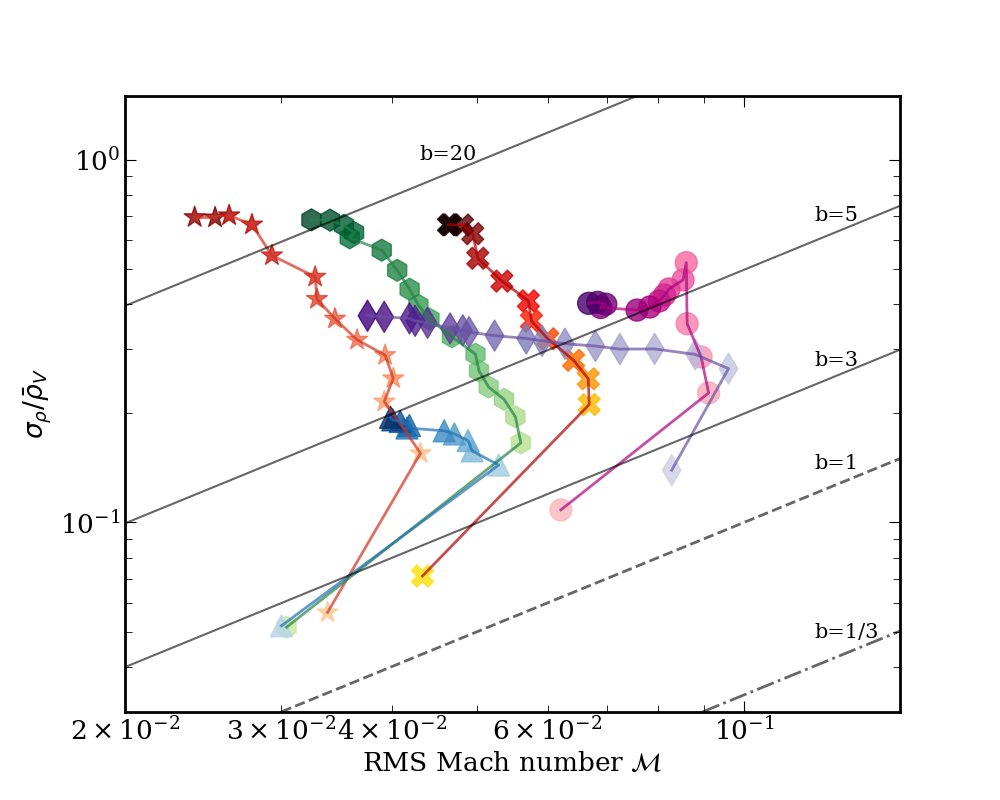}
\includegraphics[scale=0.55]{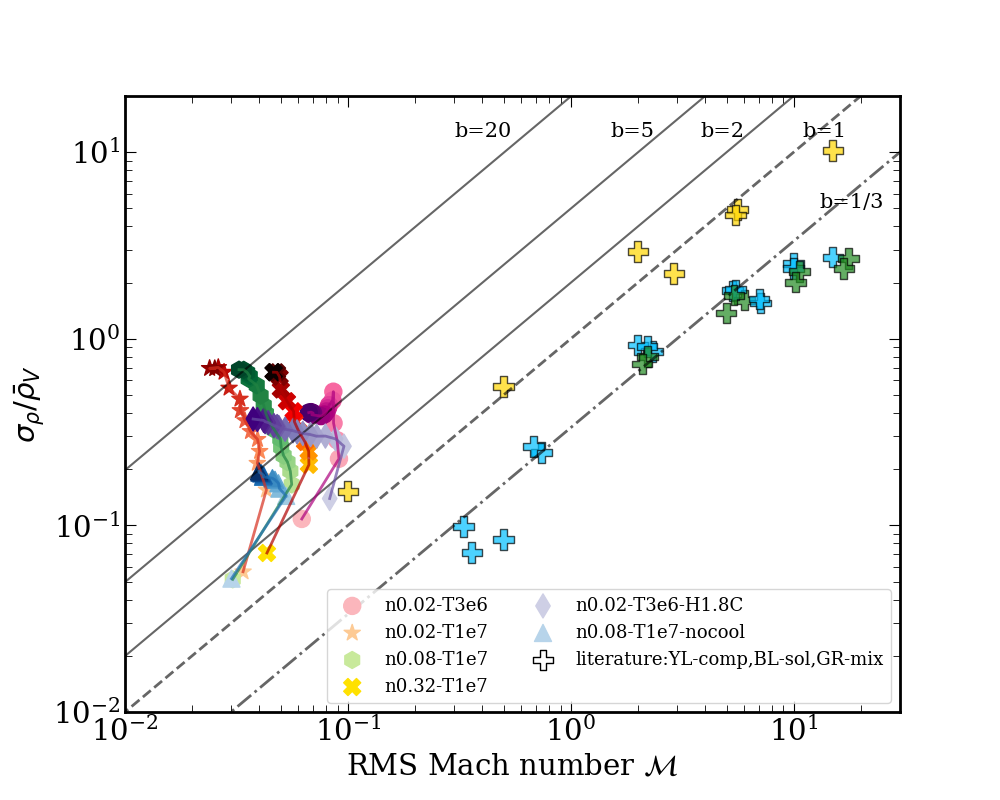}
\caption{Upper panel: Density fluctuation $\sigma_{\rho,V}/\bar{\rho}_V$ versus RMS Mach number $\mathcal{M}$ for a few sample runs. 
The shade of the symbols shows the time evolution: darker shades indicate later times. The symbols and the time sampling are the same as in Fig. \ref{f:b-t}. Contours of constant $b_\rho$ are shown by the gray diagonal lines. 
The lower panel is an extended version of the upper panel. 
Values from the literature are shown by the crosses. The yellow crosses indicate the driving scheme is purely compressional, blue shows purely solenoidal driving, and green shows a mixture of both. The literature data are compiled from \cite{hopkins15}.
 }
\label{f:sigma_M}
\end{center}
\end{figure*}

To better understand the large $b$, we show in the upper panel of Fig. \ref{f:sigma_M} the trajectories of these runs on the $\sigma_{\rho,V}/\bar{\rho}_V$ - $\mathcal{M}$ diagram. The shades of the colored symbols indicate time: later times are represented by darker shades. The time sampling is the same as in Fig. \ref{f:b-t}. Constant $b_\rho$ are indicated by the grey diagonal lines. 
The shape of the trajectories are similar for all simulations: $\sigma_{\rho,V}/\bar{\rho}_V$ is always increasing, but $\mathcal{M}$ has an initial small rise followed by a slow decline. 
Overall, the change in $\mathcal{M}$ is small, within 50\%, but $\sigma_{\rho,V}/\bar{\rho}_V$ increases by a factor of 5-30 over the course of the simulations. This indicates that the large and increasing $b$ seen in Fig. \ref{f:b-t} is due to the large and ever increasing $\sigma_{\rho,V}/\bar{\rho}_V$. (Note that the change in Mach number during its long decreasing phase is partly due to the confinement of the closed box. So the final value of $b$ can be artificially enhanced by up to a factor of 2. This is still small compared to the large increase of the density fluctuation, however. )

The subsonic Mach number has been discussed in Section \ref{sec:velocity}, which is because the blast waves driven by SNe decay into sonic waves very quickly in a hot medium. The reason for a large density fluctuation is the following: by driving blast waves, each SN punches a low-density, hot bubble in the medium; this bubble lasts long, because by the time the blast decays into a sound wave, the bubble reaches pressure equilibrium with the surroundings. If no mixing occurs, the bubble will stay permanently. Mixing processes such as turbulence and conduction are inefficient, therefore the density contrast is slow to erase. Another way to think is that SNe heat medium locally (in contrast to volumetrically), which create additional density contrast than by motions alone.
Consequently, the medium has a large density fluctuation with a small mean Mach number.

When the cooling is turned off (by comparing n0.08-T1e7-nocool and n0.08-T1e7), at later times, $\sigma_{\rho,V}/\bar{\rho}_V$ rises more slowly while $\mathcal{M}$ decreases faster. The slower increase of the density fluctuation arises because cooling instabilities cannot develop (though low-density bubbles are still being generated). The faster decline of $\mathcal{M}$ arises because the sound speed increases faster, since the mean temperature increases faster without radiative cooling. A similar pattern is seen when $H\gg C$ (by comparing n0.02-T3e6-H1.8C and n0.02-T3e6).

The lower panel of Fig. \ref{f:sigma_M} is an extended version of the upper panel, with results from other work to place our simulations in context.
The crosses in Fig. \ref{f:sigma_M} show results from the literature compiled by \cite{hopkins13c}\footnote{Table 1 of \cite{hopkins13c} listed $\sigma_s$ (their $S_{\rm{ln \rho,V}}$), which we convert into $\sigma_\rho$ using $\sigma_s^2 = ln(1+\sigma_\rho^2)$ \citep{price11,konstandin12} }.
The yellow crosses indicate the driving force of turbulence is purely compressional, the blue crosses show purely solenoidal force, and the green crosses present a mixture of both. The pure compressional driving gives $b_\rho \approx$ 1 while the pure solenoidal driving and the mixed driving give $b_\rho <$ 1. 
Our simulation results occupy the regime of low $\mathcal{M} \lesssim$ 0.1 but high $\sigma_{\rho,V}/\bar{\rho}_V \sim$ 0.1-1, which has not been found in previous simulations.

The large density fluctuations with small $\mathcal{M}$ distinguish our results from previous work. 
The difference includes several folds: (i) localized heating from SNe generates large density contrast; (ii) supersonic motions are allowed to decay, in contrast to other turbulence simulations where the driving force is maintained on the driving scale; and (iii) radiative cooling and thermal instability also contribute to a larger $b$, but this is not a major effect.

The relation between density fluctuation and the RMS Mach number has strong observational implications \citep[e.g.][]{lada94}. With $b$, we may infer the density fluctuation if the Mach number is known, and vice versa. This has been done for the molecular clouds \citep[e.g.][]{burkhart12}. Recently, the velocity dispersion of hot gas in galaxy clusters has been inferred from the spatial fluctuation of X-ray surface brightness using this correlation, adopting $b$ around unity \citep{zhuravleva14}. However, as we have found, under the impact of SNe Ia, $b$ continues to rise with time, and has values of 2-20. One may argue that this effect is on scales of individual SN bubbles, which is are well below the current angular resolution of X-ray observations (corresponding to a few kiloparsec in length scale for nearby giant elliptical galaxies). However, as we will discuss in the next section, the length scale of the density fluctuation grows with time. This is especially necessary for future X-ray missions with higher angular resolutions, such as Athena and Lynx. 

\subsection{Large-scale density structure}
\label{sec:large-scale_density}

In this section, we present an interesting phenomenon: the growth of a large-scale density structure. This is observed in all our simulations. 

\begin{figure*}
\begin{center}
\includegraphics[width=0.31\textwidth]{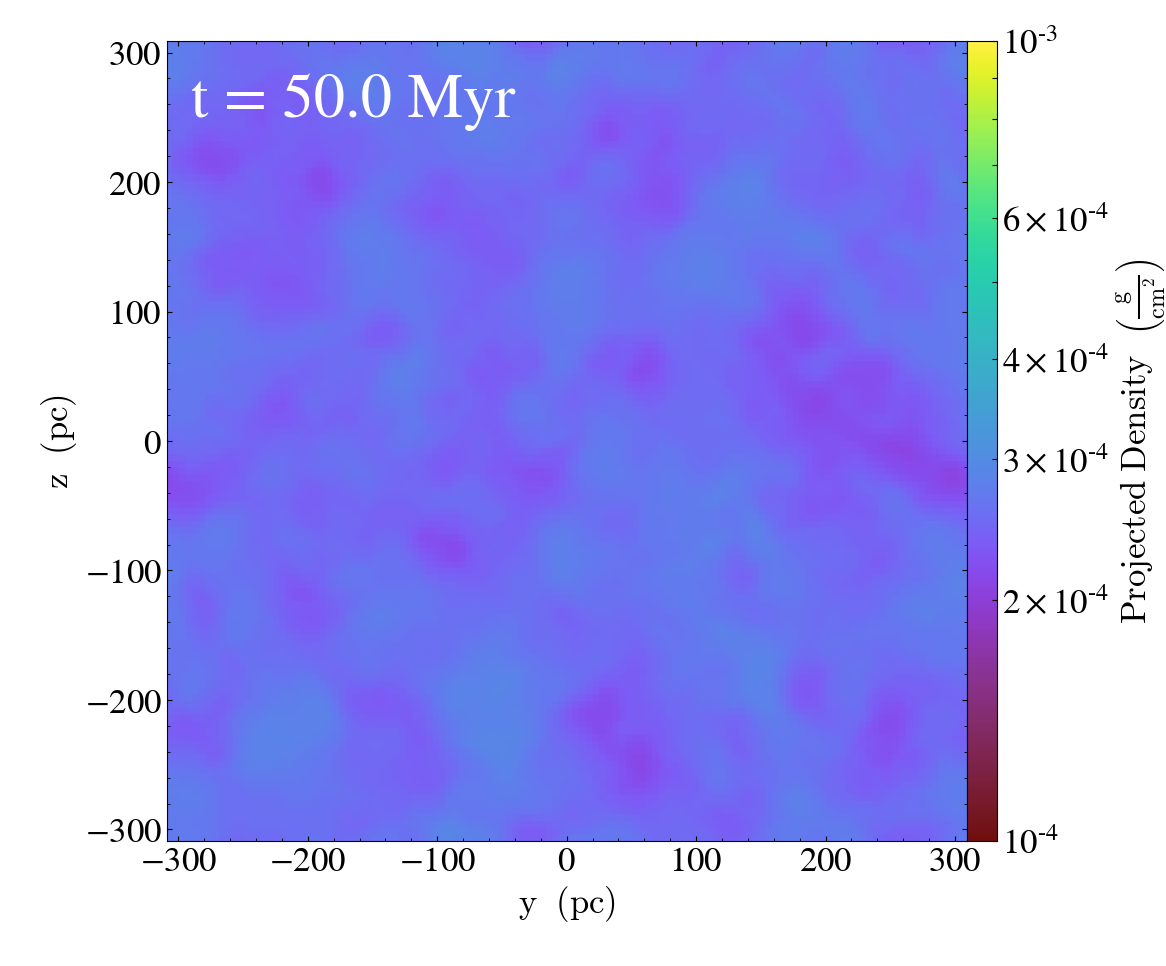}
\includegraphics[width=0.31\textwidth]{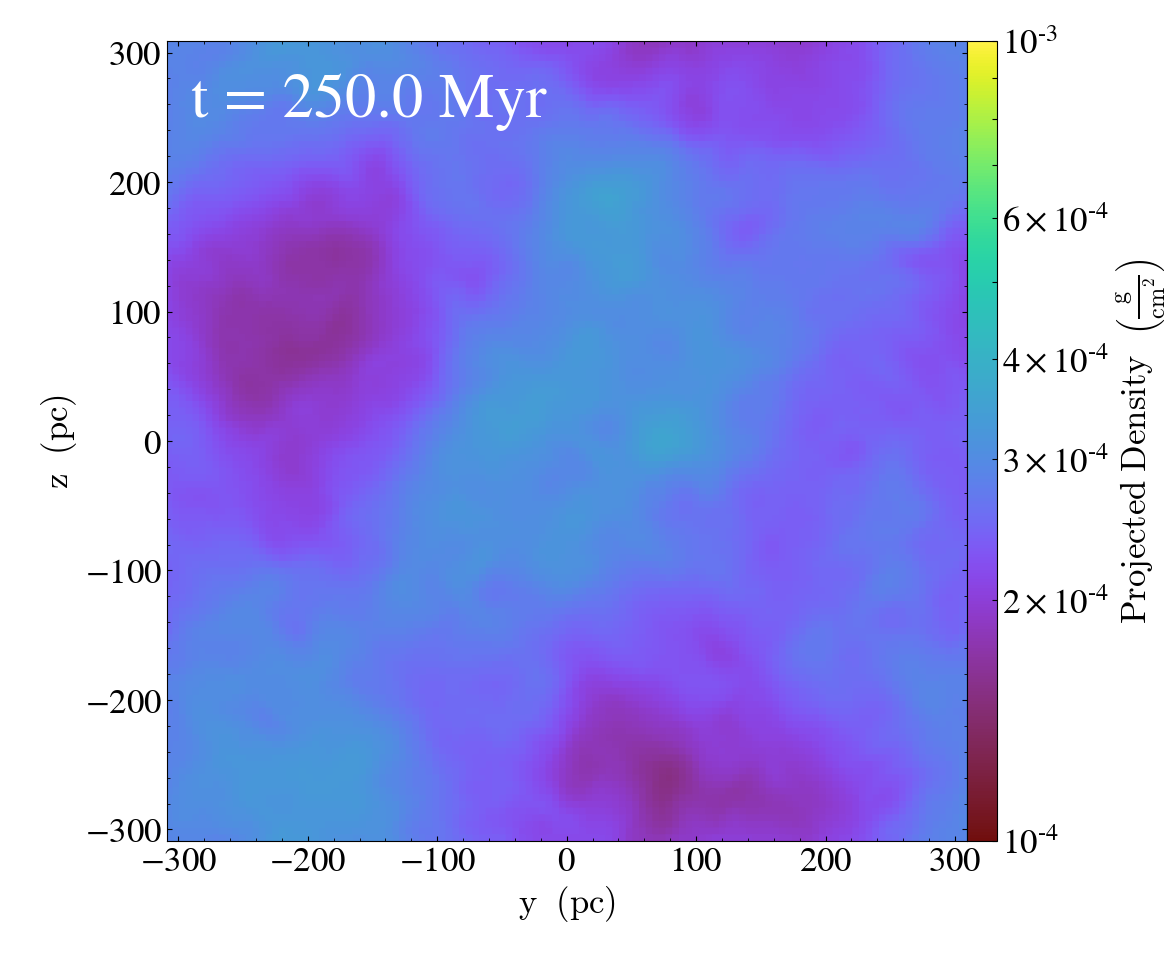}
\includegraphics[width=0.31\textwidth]{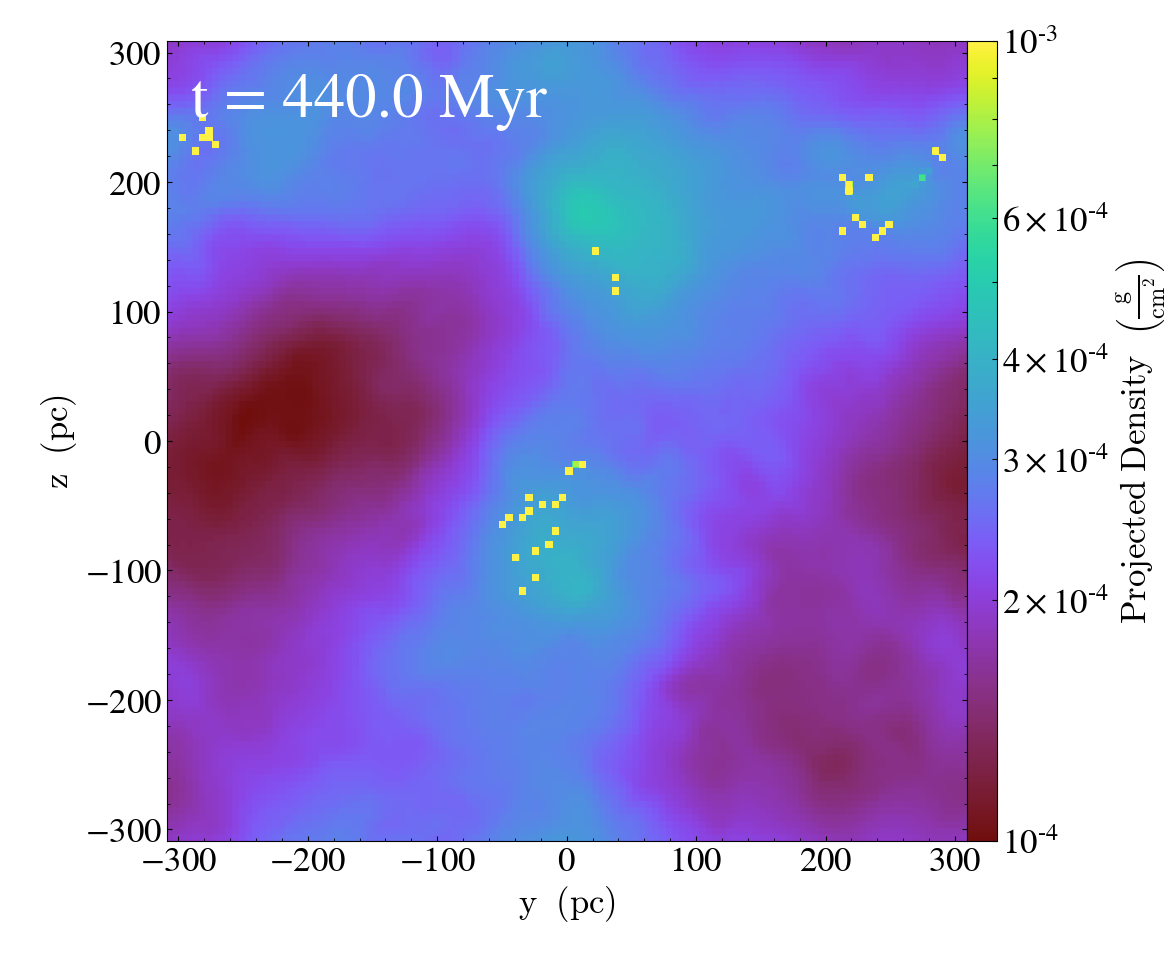}
\includegraphics[width=0.31\textwidth]{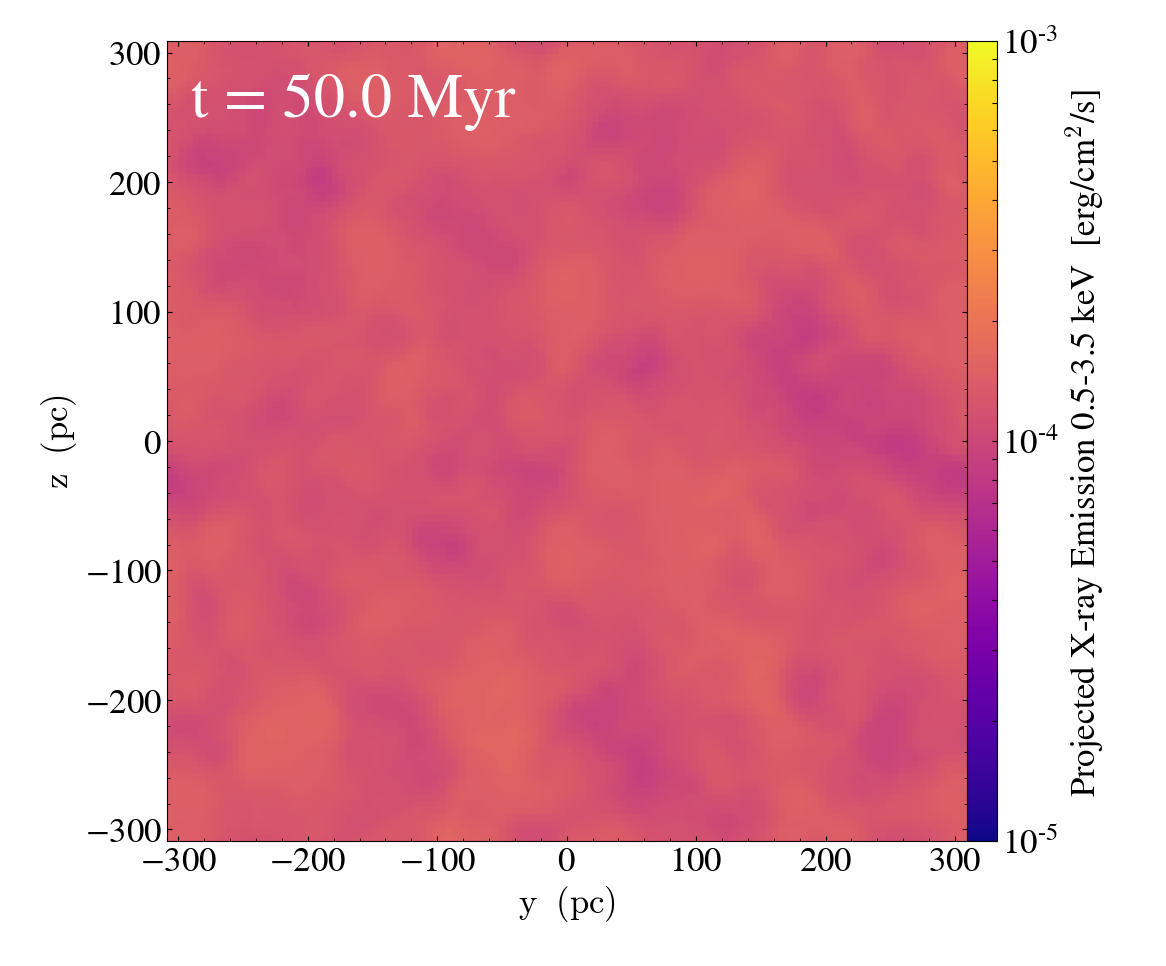}
\includegraphics[width=0.31\textwidth]{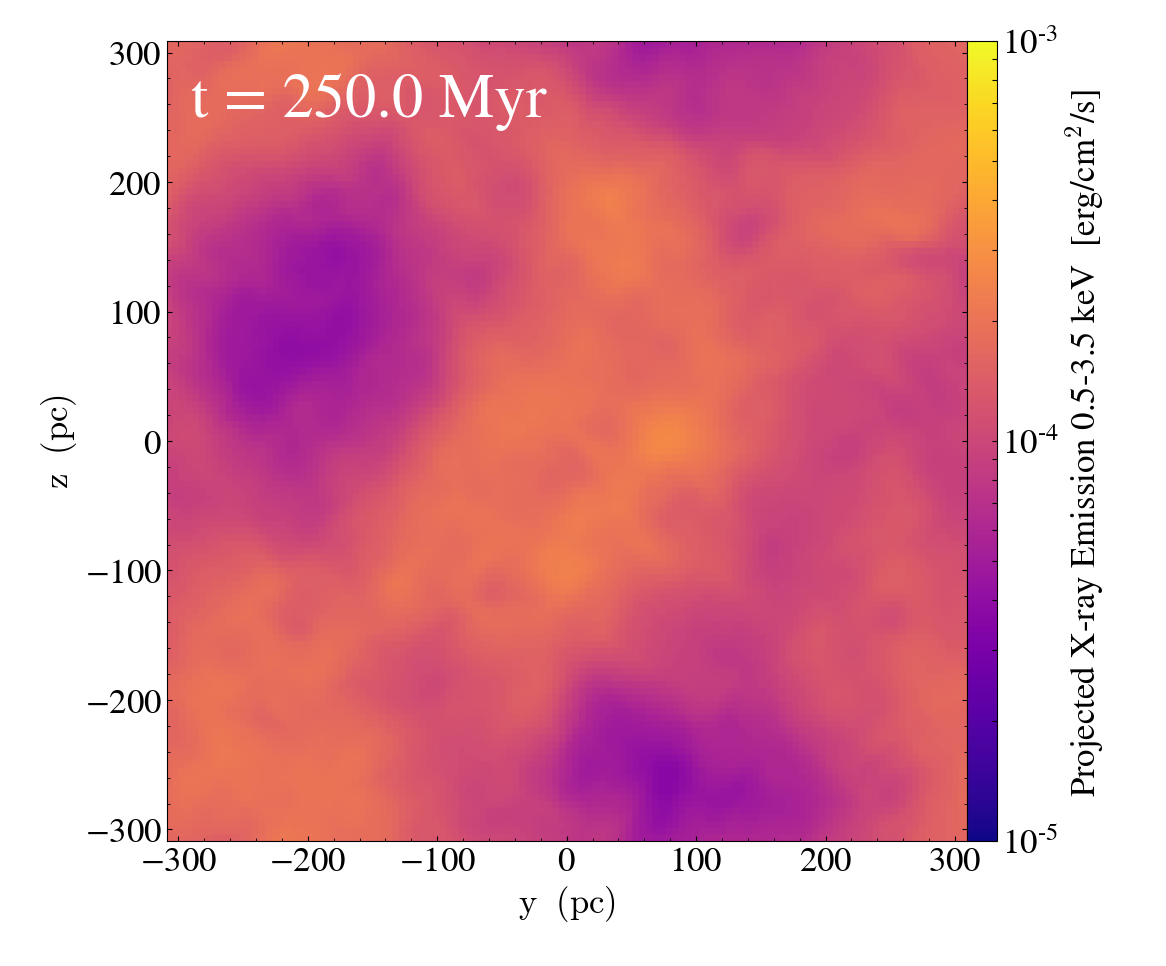}
\includegraphics[width=0.31\textwidth]{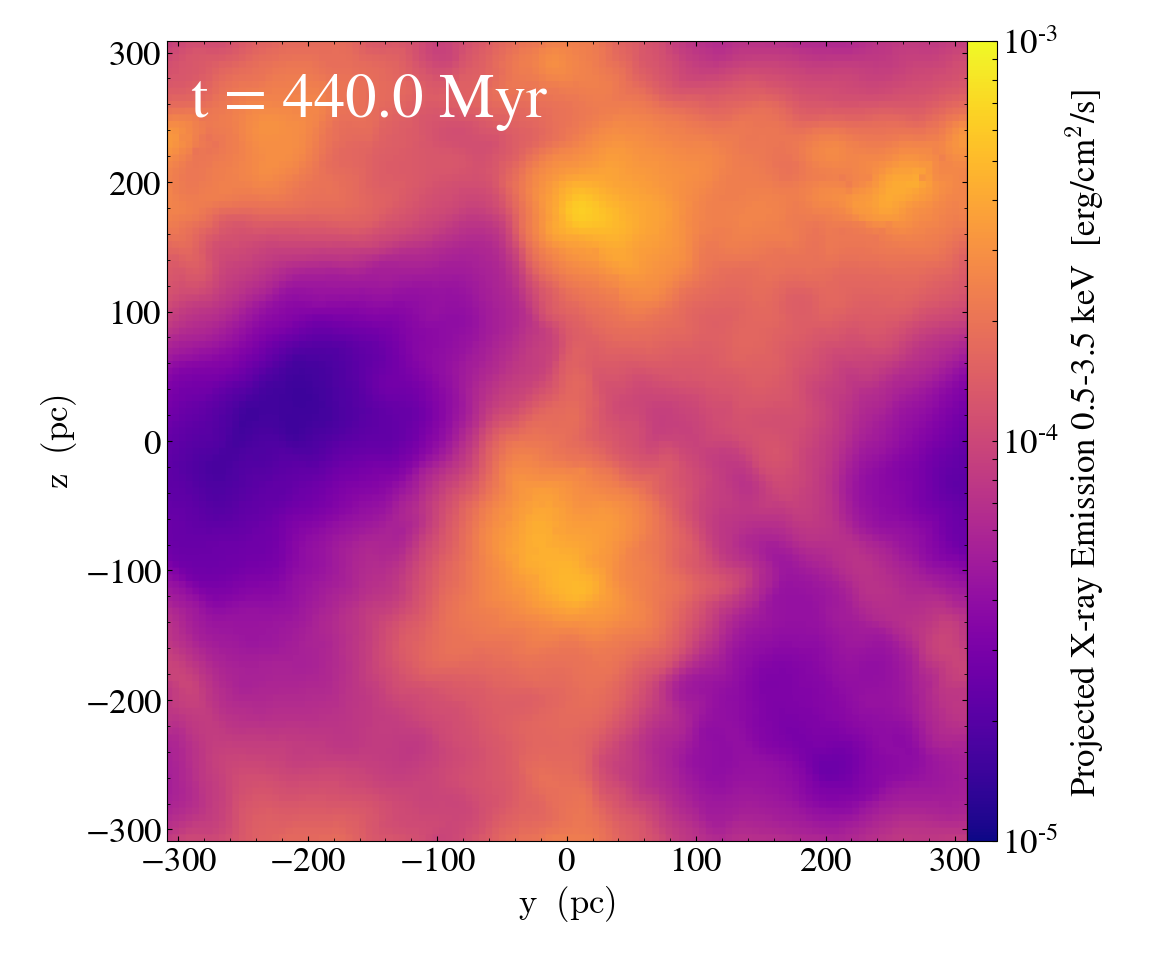}
\caption{Upper panels: Column density of n0.08-T1e7 at 50, 250, 440 Myr (0.4, 1.9, 3.0 \tc). The typical length scale of density fluctuation increases with time. Cool phase forms along the large-scale density fluctuation. The spatial fluctuation is about a factor of 10 at $t=$440 Myr. Lower panels: Projected X-ray emission (0.5-3.5 keV band) at the same times as the upper panels. There is a good spatial correlation between the X-ray emission and the density. The spatial fluctuation of X-ray is about a factor of 100 at $t=$440 Myr, larger than that of the density.
}
\label{f:projection_008_1e7K}
\end{center}
\end{figure*}

The upper panels of Fig. \ref{f:projection_008_1e7K} shows the density projections for the run n0.08-T1e7. The snapshots are taken at $t=$ 50, 250, and 440 Myr, corresponding to 0.4, 1.9, and 3.0\tc. In the last snapshot, the cool phase has formed, which manifests itself as discrete dense clumps that are saturated on the color scale. Excluding these clumps, we can see that the typical length scale of density fluctuation grows with time. In the first snapshot, the length scale is close to that of a SN bubble, which has \rsn $=$ 31 pc. In the last snapshot, however, the length scale of low-density ``bubbles'', or dense ridges, is about 200-300 pc in diameter. Considering the projection effect, the actual length scale is 350-500 pc, much larger than an individual SN bubble. Note that the cool phase forms along the dense ridges.

\begin{figure}
\begin{center}
\includegraphics[width=0.5\textwidth]{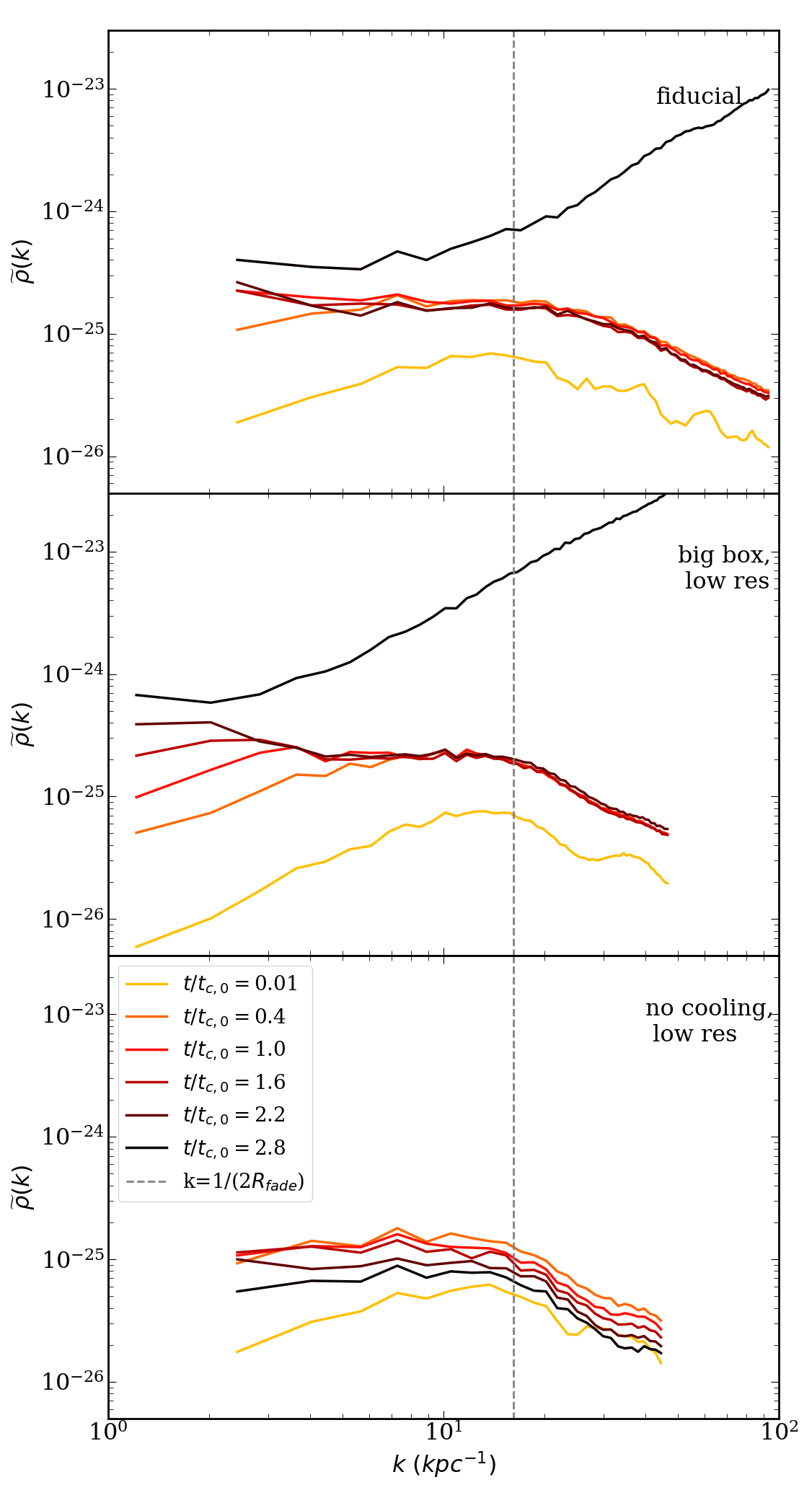}
\caption{Fourier spectrum of gas density for n0.08-T1e7. Darker colors indicate later times. The upper panel is for the fiducial run. The middle panel is for a run with a larger box and a coarser resolution. The lower panel is for a fiducial box with no cooling and a coarser resolution. With cooling, the power of density goes toward larger length scales at later times. 
}
\label{f:n008_1e7_density_spectrum}
\end{center}
\end{figure}

To be more quantitative, we plot the Fourier spectrum of gas density, $\widetilde{\rho} (k)$, in the top panel of Fig. \ref{f:n008_1e7_density_spectrum}. The vertical line indicates $k =(2$\rsn)$^{-1}$, the inverse of the size of a SN bubble. The latest snapshot is after the cool phase has formed, and the power has increased radically at large-$k$ modes. For our discussion below, we focus on the times before the cool clump formation. 
Initially, the peak of $\widetilde{\rho} (k)$ is around $k =(2$\rsn)$^{-1}$. Later, small-$k$ modes grow, corresponding to the growth of bubble size in Fig. \ref{f:projection_008_1e7K}. At $t\geqslant $\tc, $\widetilde{\rho} (k)$ is largest at the smallest $k$. This confirms the growth of the large-scale density structure. 
In contrast, the short-wavelength modes, with $k\gtrsim$ 15 kpc$^{-1}$, reach a steady state after an initial growth. 

The effect on the Fourier spectrum is more prominent when the box size is larger. We show in the second panel of Fig. \ref{f:n008_1e7_density_spectrum} a run with the box size twice larger (while the resolution is 2 times coarser to save the computing cost\footnote{We have checked that the resolution itself does not affect the result.}). It is consistent with the fiducial run for the range of $k$ where the two simulations overlap. The power on the smallest $k$ continues to grow. Note that this shift of the peak in the Fourier spectrum only applies to the density. The spectrum of the kinetic energy remains unchanged over time, with a peak at $k = (2$\rsn)$^{-1}$. When we turn off cooling (lower panel of Fig. \ref{f:n008_1e7_density_spectrum}), the small-$k$ modes do not show a continual growth.

This growth in density structure is universally seen in our simulations. We attribute the growth to the following reason: thermal instability grows on all scales, but it is suppressed under the scale of turbulent diffusion, which evens out the density fluctuation. The turbulent diffusion acts on scales larger than the driving scale $l_{\rm{drive}}$, with a diffusivity $D \approx l_{\rm{drive}} v_{\rm{turb}} = 2$\rsn$v_{\rm{sol}}$. Therefore, the scales at which the density power peaks, $L$, increases with the square root of time, 
\begin{equation}
    \frac{L^2}{t} = D  \approx 2R_{\rm{fade}} v_{\rm{sol}}.
\end{equation}
Thus,
\begin{equation}
\begin{split}
   \frac{L(t)}{2R_{\rm{fade}}} 
    & =\left(\frac{v_{\rm{sol}} t_{\rm{c,0}} }{R_{\rm{fade}}} \right)^{1/2} \left(\frac{t}{t_{c,0}}\right)^{1/2} \\
    & =\left(\frac{c_s t_{\rm{fade}}}{R_{\rm{fade}}} \mathcal{M} \frac{v_{\rm{sol}}}{v_{\rm{RMS}}} \frac{t_{c,0}}{t_{\rm{fade}}}   \right)^{1/2} \left( \frac{t}{t_{c,0}}  \right)^{1/2} ,
\end{split}
\end{equation}
Here $t_{\rm{fade}} c_s = 0.36R_{\rm{fade}}$, where the coefficient is obtained from the Sedov solution. Plugging in the numbers, we obtain
\begin{equation}
\begin{split}
      \frac{L(t)}{2R_{\rm{fade}}} =
  & \  6.5  \left( \frac{\mathcal{M}}{0.05} \right) ^{1/2} \left( \frac{v_{\rm{sol}}/v_{\rm{RMS}} }{0.2} \right) ^{1/2} \\ & \left(\frac{t_{c,0}/t_{\rm{fade}}}{5500}\right)  ^{1/2}   \left(\frac{t}{3t_{c,0}}\right)^{1/2},
\end{split}
\end{equation}
where the values are for this run (although all the simulations have values of the same order of magnitude). This means that at $t=3$\tc, $L\approx$ 13\rsn $\approx$ 400 pc, which is consistent with the bubble size seen in Fig. \ref{f:projection_008_1e7K}.

This spatial density fluctuation will translate into the fluctuation of X-ray emission. The lower panels of Fig. \ref{f:projection_008_1e7K} show the projected X-ray emission for the energy range 0.5-3.5 keV. There is a good spatial correlation between the X-ray emission and the density: the denser regions are brighter in X-rays, and vice versa. At later times, the variation of X-ray emission becomes larger, similar to that of the density. Note that the amplitude of the X-ray fluctuation is larger than that of the density. At $t=440$ Myr, the spatial variation of X-ray emission spans two orders of magnitude, whereas that of the density spans only one order of magnitude (here we only refer to the hot gas, not to the dense cool clumps, since the latter do not emit X-rays). This is understandable since the emission from two-body collisions scales with density squared. 
This density fluctuation, or ``clumping'' effect, is very important when X-ray luminosity is used to infer the underlying mass of the hot gas. Also, as discussed in the previous section, the bubbles and ridges may contribute to the fluctuation of the X-ray surface brightness of hot gas in early-type galaxies and galaxy clusters. 

As discussed in Paper I, the inhomogeneity of the density leads to buoyancy in a gravitational potential-- the low-density patches tend to rise and the higher-density ones tend to sink. It will be very interesting to see how the density fluctuation respond in a stratified medium. We postpone this to future studies.

Note that our experiments are designed to examine the evolution of gas properties over a specific time scale, i.e., the cooling time of the hot medium. Over this time, the kinematics of the gas reach a steady state quickly after the simulation starts, as discussed in Section \ref{sec:velocity}; in contrast, quantities related to the thermal state of the gas, including the density distribution, thermal energy, and the parameter $b$, still evolve at the end of the simulation. (The development of the thermal instability means that thermal properties will not reach steady states over the thermal timescale.) For these still-evolving quantities, we have quantified them based on their change over the cooling time. Since the simulation time is much shorter than the age of the galaxy and the experiments are idealized in nature, the exact values of these still-evolving quantities found in simulations may not reflect those in reality. However, for the conditions we have assumed, it is inevitable that gas heated by SNe Ia evolves away from uniformity and undergoes overheating, therefore these small-scale effects should be considered when observations are interpreted, as well as in coarse-resolution simulations. For example, when information is extracted from observations, a gas model with a broad (log-normal) density/temperature distribution can be used; for cosmological simulations, we may attempt to compensate in some way for the missed overheating effect of the unresolved SNe Ia. The long-term and large-scale impact of SNe through these small-scale processes is beyond the scope of this paper and needs to be investigated in a global simulation.

\section{Summary}

In this paper we analyze a series of simulations with resolved SNe Ia feedback in the hot ISM of elliptical galaxies. We examine the energetics and turbulence structure of the medium. 
The localized and distributed feedback from SNe Ia leaves distinct features in the hot medium.
The main conclusions are the following:

(1) When SN remnants are resolved, the energy evolution of the ISM patch behaves very differently from when SNe are treated as a volumetric heating term. The net heating rate appears significantly higher than $H-C$ for almost all conditions (Figs. 1, 2). This is due to the rarefaction effect of the SN-driven blast waves, which reduces the density of a significant fraction of the medium, thus lowering the mean cooling rate (Fig. 3). This effect is missing when SNe Ia are treated as subgrid heating in cosmological simulations. 

(2) The RMS velocity of gas is 20-50 km s$^{-1}$ on a driving scale of tens of pc. This is significantly lower than the sound speed of the hot gas in elliptical galaxies, which is 200-500 km s$^{-1}$. The RMS Mach number is thus small, $\approx$ 0.05-0.15 (Fig. 5).

(3) The velocity field of the medium is dominated by the compressional component. The magnitude of the compressional component is 3-8 times that of the solenoidal (Fig. 6).

(4) The empirically measured turbulent mixing time from Paper I, $t_d$, is found to be very close to the theoretical estimate of the turbulent cascade time, $t_{\rm{cas}} \sim 2R_{\rm{SN}}/v_{\rm{sol}}$ (Fig. 7).

(5) The density distribution is close to log-normal (Fig. 8), especially at late times. 

(6) The value of $b$ from Eq. \ref{eq:rho_bM}, which is the ratio between the spatial fluctuation of the gas density and the RMS Mach number, is 2-20 in our simulations (Fig. 9). This is much higher than what has been found in previous work, where $b\lesssim$ 1 (Fig. 10). This is mainly because the localized SN heating generates an additional large density contrast.
 
(7) The density of the ISM shows the formation of large-scale structure over time. The size of low-density bubbles and high-density ridges continues to grow. This is likely due to a combined effect of thermal instability and mixing through SNe-driven turbulence (Fig. 11, 12). \\

\section{Concluding Remarks}

In a series of two papers, we investigate the impact of SNe Ia on the hot ISM, which is typical in quiescent galaxies. The medium exhibits a rich and unique spectrum of features. Some of the main features seem counterintuitive at first. For example,
(i) the gas is overheated while the cool phase is allowed to form;
(ii) SNe Ia drive blast waves but the overall Mach number of the medium is quite small, $\lesssim$ 0.1;
(iii) SNe Ia drive turbulence from small scales (tens of pc), but the gas show density structures on large scales ($\gtrsim$ several hundred pc). 

The interesting phenomena come from the interplay between SNe feedback and the hot medium -- more specifically, the heating and motions caused by many SNe on small scales, and thermal instability that the hot medium is prone to. The physics together make the medium inhomogenized and turbulent. These effects have generally not been included in the studies of quiescent galaxies. Future experiments incorporating more complete physics, such as gravity and stratification, will provide a deeper understanding of the dynamics and thermodynamics of the ISM under SNe Ia, and how they affect galaxy evolution.

\section*{Acknowledgement}
We thank the referee for helpful comments which improve the clarity of the paper.
We thank members of the SMAUG collaboration for useful discussions. ML thanks Jeremiah Ostriker and Feng Yuan for helpful discussions, and John Forbes for making probability distribution plots \footnote{\url{https://arxiv.org/abs/2003.14327}}. Computations were performed using the publicly-available Enzo code, which is the product of a collaborative effort of many independent scientists from numerous institutions around the world. Their commitment to open science has helped make this work possible. Data analysis and visualization are partly done using the \textsf{yt} project \citep{turk11}. The simulations are performed on the Rusty cluster of the Simons Foundation and the XSEDE clusters supported by NSF. We thank the Scientific Computing Core of the Simons Foundation for their technical support.
We acknowledge financial support from NSF (grant AST-1615955, OAC-1835509 to GB, AST-1715070 to EQ), NASA (grant NNX15AB20G to GB), and the Simons Foundation (grant 510940 to ECO, 528306 and a Simons Investigator Award to EQ).

\vspace{0.2in}

\bibliographystyle{aasjournal}

\end{CJK*}
\end{document}